\documentclass{article}

\usepackage{arxiv}

\usepackage[utf8]{inputenc} 
\usepackage[T1]{fontenc}    
\usepackage{hyperref}       
\usepackage{url}            
\usepackage{booktabs}       
\usepackage{amsfonts}       
\usepackage{nicefrac}       
\usepackage{microtype}      
\usepackage{graphicx}
\usepackage{natbib}
\usepackage{doi}
\usepackage{enumerate}
\usepackage{multirow}
\usepackage{float}
\usepackage{subfigure}
\usepackage{url}
\usepackage{booktabs}
\usepackage{paralist}
\usepackage[normalem]{ulem}
\useunder{\uline}{\ul}{}
\usepackage{lscape}

\title{Towards Web of Things Middleware: A Systematic Review}

\author{ 
	{\hspace{1mm}Habib Larian}\\
	Faculty of Computer Engineering\\
	Najafabad Branch, Islamic Azad University\\
	Isfahan, Iran\\
	\texttt{h.larian@yahoo.com} \\
	\And
	{\hspace{1mm}Ali Larian}\\
	Department of Electrical and Computer Engineering\\
	Isfahan University of Technology\\
	Isfahan, Iran\\
	\texttt{a.larian@ec.iut.ac.ir} \\
	\And
	 {\hspace{1mm}Mahdi Sharifi}\\
	 Faculty of Computer Engineering\\
	 Najafabad Branch, Islamic Azad University\\
	 Najafabad, Iran\\
	 \texttt{m.sharifi@pco.iaun.ac.ir} \\
	\And
	{\hspace{1mm}Homa Movahednejad}\\
	Faculty of Computer Engineering\\
	Najafabad Branch, Islamic Azad University\\
	Najafabad, Iran\\
	\texttt{h.movahed@pco.iaun.ac.ir} \\
	}

\date{}

\renewcommand{\shorttitle}{Towards Web of Things Middleware}

\hypersetup{
pdftitle={Towards Web of Things Middleware: A Systematic Review},
pdfsubject={},
pdfauthor={Habib Larian, Ali Larian},
pdfkeywords={Web of Things, Internet of Things, WoT Architecture, IoT Middleware, WoT Platforms},
}

\begin{document}
\maketitle

\begin{abstract}
	Advancements of the Web technology provide this opportunity for Internet of Things (IoT) to take steps towards Web of Things (WoT). By increasing trend of reusing Web techniques to create a monolithic environment to control, monitor, and compose the smart objects, a mature WoT architecture is finally emerged in four layers to be a solution for IoT-middleware. Although WoT architecture facilitates addressing requirements of IoT in architectural or service aspects, but the effectiveness of this solution is indeterminate to meet IoT-middleware objectives. The most surveys and related reviews in this field just investigate IoT-middleware and WoT separately and thereby, report some new technologies or protocols on various middlewares or WoT models. In this paper a comprehensive survey is proposed on common area of IoT and WoT disciplines by leveraging Systematic Literature Review (SLR) as research methodology. This survey classifies variant types of IoT-middleware and WoT architecture to specifies their requirements and characteristics, respectively. Hence, WoT requirements could be categorized by comparing and analyzing IoT-middleware requirements and WoT characteristics. This research heavily reviews existing academic and industrial contributions to select potential platforms (or frameworks) and assess them against proposed WoT requirements. As a result of this survey, strengths and weaknesses of WoT architecture, as a IoT-middleware, are presented. Finally, this research attempts to open new horizon for the WoT architecture to enable researchers to dig role of WoT technologies in the IoT.
\end{abstract}


\section{Introduction}
By progressing in communication technologies, physical objects become smart by equipping to network communication chips in order to enable users to control and manage them remotely. By associating these smart objects together, IoT provides an opportunity to create various platforms in industrial or daily life applications. Smart objects are turned to resources of information when connect to the Internet. Hence, the Internet not only connects people to resources of information, but also links these resources together. This level of communications flows the information traffic via the Internet highway \cite{RN131,RN686}.

Emerging IoT accelerates information flow in the world. Advancements of IoT try to harness this flow by forming a functional ecosystem and directing information. This ecosystem is built on integration of interactive smart sub-systems called cyber-physical systems \cite{RN659}. Cyber-physical systems are the main information resources to generate crucial data. It is a challenging task to integrate these cyber-physical systems to create a monolithic smart environment by own particular standards or protocols. Therefore, reusing a conventional technology such as Web is an efficient and low-cost solution to address heterogeneity of technologies in such coherent ecosystems. WoT as a new concept has emerged to reconcile World Wide Web technologies to the cyber-physical systems. Practically, WoT transforms cyber-physical systems from being traditional IoT into Web resources where information is generated and exposed within services \cite{RN18}.

Traditional IoT systems were developed to control and monitor smart environments wherein several cyber-physical systems integrated together and build an interactive IoT platform based on Machine to Machine (M2M) communication model. It means, a specific platform had to be developed for each smart environment. The basic idea of IoT-middleware is emerged to eliminate M2M communications and provides a framework to make a seamless interoperation between cyber-physical systems and end-users \cite{RN303}. At the present, variant types of IoT-middleware are developed to satisfy certain IoT ecosystem requirements. WoT leverages service-oriented middleware as a heart of its architecture to handle the complex requirements of IoT ecosystems. WoT architecture relies on two types of service-oriented model including Service-oriented Architecture (SoA) and Resource-oriented Architecture (RoA) \cite{RN63,RN682}. A basic WoT architecture is designed as an IoT framework by Guinard \cite{RN657} in four layers across the Web technologies.

\begin{table}[t]
	\centering
	\caption{Research Questions}
	\label{tab:my-table}
	\resizebox{\textwidth}{!}{%
		\begin{tabular}{@{}cll@{}}
			\toprule
			\multicolumn{1}{l}{\textbf{Level}} &
			\multicolumn{1}{c}{\textbf{RQ code}} &
			\multicolumn{1}{c}{\textbf{RQ Statement}} \\ \midrule
			\multirow{7}{*}{\textbf{Ex\#0}} &
			\textbf{L0.RQ} &
			\textbf{\begin{tabular}[c]{@{}l@{}}How many requirements of the IoT-middleware are covered by characteristics of the Web of Things architecture?\end{tabular}} \\
			& L0.RS1.SRQ01 & What is the IoT-middleware?                      \\
			& L0.RS1.SRQ02 & How many types of IoT-middleware exist?         \\
			& L0.RS1.SRQ03 & What are the requirements of the IoT-middleware? \\ \cline{2-3} 
			& L0.RS2.SRQ01 & What is the Web of Things?                       \\
			& L0.RS2.SRQ02 & How many types of WoT architecture exist?        \\
			& L0.RS2.SRQ03 & What are the characteristics of WoT architecture? \\ \midrule
			\multirow{2}{*}{\textbf{Ex\#1}} &
			\textbf{L1.RQ} &
			\textbf{\begin{tabular}[c]{@{}l@{}}How many WoT requirements are supported by WoT   platforms?\end{tabular}} \\
			&
			L1.RS1.SRQ01 &
			\begin{tabular}[c]{@{}l@{}}Which IoT-middleware platforms could be adapted to WoT architecture? (stated-of-the-art)\end{tabular} \\ \bottomrule
		\end{tabular}%
	}
\end{table}

The most existing surveys on IoT-middleware and WoT only focused on investigating new technologies or communication protocols on each topic separately. For example, Zeng et al. \cite{RN107} present a review on types of WoT architecture and some key enabling technologies of WoT. Atzori et al. \cite{RN131} investigate IoT from three main visions and subsequently present a review on IoT architecture, its enabling technologies and applications. Khoi tran et al. \cite{RN239} propose state-of-the-art WoT search engines and develop a framework to assess them. The authors in \cite{RN145} classify IoT-middlewares in various models and offer a set of requirements by considering IoT characteristics and eventually, investigate middleware solutions against these requirements.

Although the WoT architecture is a solution to integrate Web and cyber-physical systems together, however the effectiveness of this solution is not yet properly demonstrated from IoT-middleware viewpoint. This systematic literature presents an investigation on both IoT-middleware and WoT to realize some requirements of WoT ecosystems in order to evaluate WoT platforms. The main contributions of this article are summarized as follows.

\begin{itemize}
	
	\item Leveraging SLR to make an efficient research strategy in order to ask right questions and design a proper protocol to explore research area of IoT and WoT to address these questions (as indicated in table 1).
	\item Reviewing IoT-middleware field, proposing a clear definition for IoT-middleware and classifying its variant types according to their abstraction levels and subsequently, identifying and categorizing IoT-middleware requirements in two service and architecture sub-sets.
	\item Reviewing WoT architecture in order to find out a clear definition and its various types based on architectural models and thereby, realizing WoT architecture characteristics according to the main elements of development process.
	\item Mapping IoT-middleware requirements to WoT ones according to compared IoT-middleware requirements with WoT characteristics and thereby, assort these requirements in two major groups as WoT-service and WoT-architectural.  
	\item Finding 17 potential WoT platforms, evaluating them based on the WoT requirements, finally reporting the strengths and weaknesses of WoT architecture in supporting of these requirements.
	
\end{itemize}

The rest of this paper is organized as follows. Sections 2 presents the employed research methodology in this paper. The third Section reviews IoT-middleware and its relevant issues in order to address IoT-middleware SLR questions. Section 4 focuses on WoT architecture to answer its questions, followed by section 5 that states analysis on WoT and IoT-middleware concepts to address main question of this research by discovering WoT requirements. The sixth section reviews 17 WoT platforms and assesses them against WoT requirements. Finally the work is concluded with reporting the strengths and weaknesses of WoT architecture.

\section{Methodology}
\subsection{Systematic Review}

\begin{figure}[b]
	\centering
	\subfigure[]{\includegraphics[width=.35\textwidth]{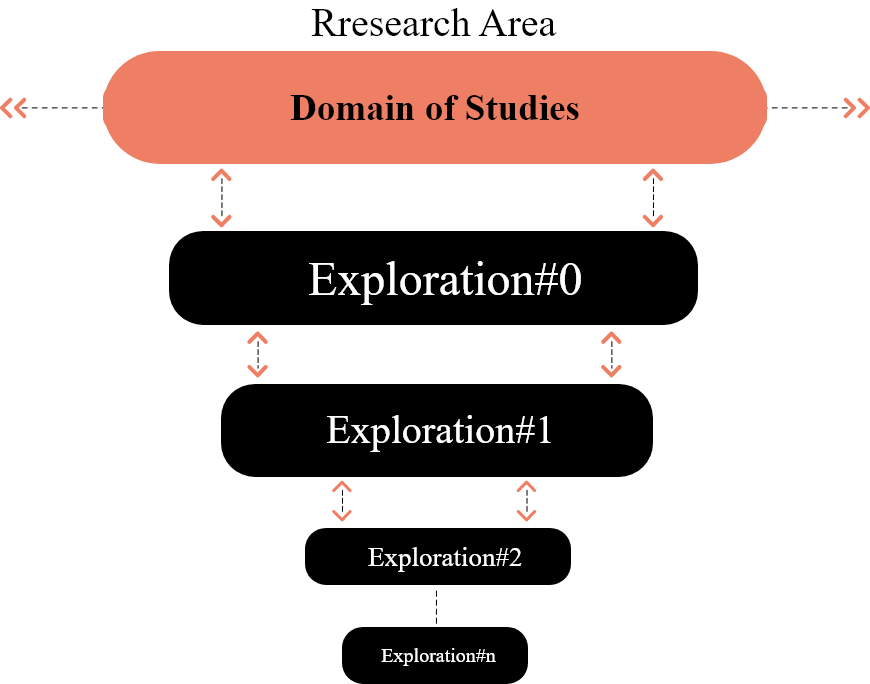}}
	\caption{Research Protocol}
\end{figure}

A Systematic Literature Review (SLR) is a research methodology that identifies, evaluates, and analyzes, all published studies relevant to a certain Research Question (RQ) or a Research Topic. In SLR, researchers synthesize all available works fairly in order to identify and report both studies either support or not support their hypotheses. This methodology is performed in three main phases as planning, conducting, and reporting. Planning include identification of need for a review, specifying the RQ(s), and developing a review protocol. In conducting, researchers start to investigate primary studies, extracting data and finally synthesis them. Lastly, researchers format their results and report them as a thorough survey in reporting \cite{RN1}.

\subsection{Planning the Review}
\subsubsection{Research Targets}
Four major targets are declared for this research as follows.
\begin{enumerate}[I.]
	\item Assessing the WoT architecture as a potential solution to address IoT-middleware is the main target of this paper.
	\item Determining evaluation criteria to appraise existing WoT platforms or frameworks.
	\item Selecting the most appropriate WoT platforms and assessing them based on the identified evaluation criteria.
	\item Demonstrating weaknesses and strengths of WoT architecture in fulfilling requirements of IoT middleware is the final target of this survey.
\end{enumerate}
\subsubsection{The need for SLR}

The major target of this survey is to evaluate the potential WoT architecture to meet needs of IoT ecosystems. This target specifies a problem to answer unaddressed questions. However, this problem consists of distinct and inseparable fields (i.e., IoT and WoT) together. Hence, the presented paper has been leveraged systematic methodology to distinguish this problem and focus on main research targets. In practice, systematic methodology arranges this problem as a \textit{Domain of Studies} (DoS), which is separable by defining RQ(s) in hierarchical levels.

\subsubsection{Protocol}
\begin{table}[b]
	\centering
	\caption{Research Criteria}
	\label{tab:my-table}
	\resizebox{0.7\textwidth}{!}{%
		\begin{tabular}{@{}ll@{}}
			\toprule
			\multicolumn{1}{c}{\textbf{Exploration\#0}} & \multicolumn{1}{c}{\textbf{Exploration\#1}} \\ \midrule
			\multicolumn{2}{c}{\textbf{Inclusion criteria}}            \\ \midrule
			WoT architecture                   & WoT platform          \\
			Types of WoT architecture          & WoT framework         \\
			WoT platform                       & WoT middleware        \\
			IoT platform                       &                        \\
			Characteristics of WoT architecture & WoT architecture                            \\
			IoT middlewares                    &                       \\
			Requirements of the IoT middleware &                       \\
			Characteristics of IoT middlewares &                       \\ \midrule
			\multicolumn{2}{c}{\textbf{Exclusion criteria}}            \\ \midrule
			IoT application                    & Web services platform \\
		    Web services                       & Wisdom Web of Things  \\
			Web applications                   &                       \\ \bottomrule
		\end{tabular}%
	}
\end{table}

To arrange the DoS and separate research topics, this paper have proposed a research protocol based on the systematic methodology which limits the DoS into distinct research areas. Every research area explores DoS from a certain viewpoint. Each topic of research area is distinguished by defining the particular RQ. Research areas turn to Exploration levels by the defined RQ(s). Exploration levels are operational units of research process. Every Exploration level has own specific RQ investigated independently to find out the answer. Figure 1.a depicts how Exploration levels focus on a certain part of DoS.

Each Exploration level must be divided into sub-areas by defining sub-RQs (SRQ) to achieve better research topics. Sub-areas are described as practical research units, so-called Research Strings (RS). Every RS contains a set of keywords and strings to be used in search of primary studies.

To select potential studies unbiasedly, selection criteria must be specified in Exploration levels. The selection criteria consist of practical filters (e.g., publisher, publication date, content type, and etc.) and inclusion and exclusion criteria. Practical filters indicate some filters applied on digital libraries due the search process to restrict research results. Inclusion criteria refer to supporting criteria with respect to each RQ whereas exclusion criteria determine a set of criteria which bias the research process. Table 2 indicates research criteria of this survey in both Exploration levels. In addition, entire Exploration levels execute on a set of repositories (e.g., Web of Science or Scopus) which are used in primary studies search.

The proposed protocol is designed in 10 steps as follows:
\begin{description}
	\item [Step1. Obtaining the Research Exploration level:] 
	A new Exploration level is defined on the DoS at the beginning of research process or after finding the answer of RQ at the prior Exploration level.
	\item [Step2. Specifying the RQ:] 
	The RQ must be specified for a new Exploration level to obtain targets of research process.
	\item [step3. Defining sub-RQ(s):]
	An Exploration level could comprise sub-areas. These sub-areas must be separated by defining SRQ(s).
	\item [Step4. Describing RS:]
	Sub-areas are determined by describing RS(s). Every RS contains a set of keywords and strings that used in research process.
	\item [Step5. Defining selection criteria:]
	Selection criteria must be defined to assess existing studies. They are composing of inclusion and exclusion criteria.
	\item [step6. Selecting digital repositories:]
	A set of digital repositories must be specified with regard to research topics.
	\item [Step7. Searching papers:]
	A search algorithm is executed on predefined set of digital repositories to find potential papers.
	\item [Step8. Selecting papers:]
	To reduce bias in the selection process, a procedure should be run to select potential papers.
	\item [Step9. Extracting data:]
	By selecting appropriate studies, data must be extracted to find out the answer of SRQ(s).
	\item [Step10. Synthesizing data:]
	Extracted data must be synthesized and analyzed to address main RQ(s). If research process reaches the last Exploration level, it must be started to report the results, otherwise, next Exploration level should be defined and started from step one.
\end{description}

\subsubsection{Search Algorithm}
This algorithm is developed to search a certain RS by own strings and keywords and find the best suited papers. As depicted in figure 1.b, this algorithm is executed as follow steps.
\begin{enumerate}
	\item Starting the search algorithm.
	\item Entering specific RS strings and keywords for searching process.
	\item Running a search process on each digital repository (Scopus, Web of Science, IEEE xplore) separately by applying practical filters. 
	\item Outputting \#m number of papers as total results of searching process on set of digital repositories.
	\item Running primary\_Screen procedure to select primary studies for RS.
	\item Outputting \#n number papers (\#n <= \#m) as results of primary\_Screen procedure to add in RS paper list.
	\item Running Screen procedure to select the best suited studies to enter Extracting data step as next step of research protocol.
	\item Obtaining \#q number of papers as results of Screen procedure to add in Exploration\# selected list.
	\item Ending the search algorithm.
\end{enumerate}

\begin{figure}[h]
	\centering
	\subfigure[]{\includegraphics[width=.9\textwidth]{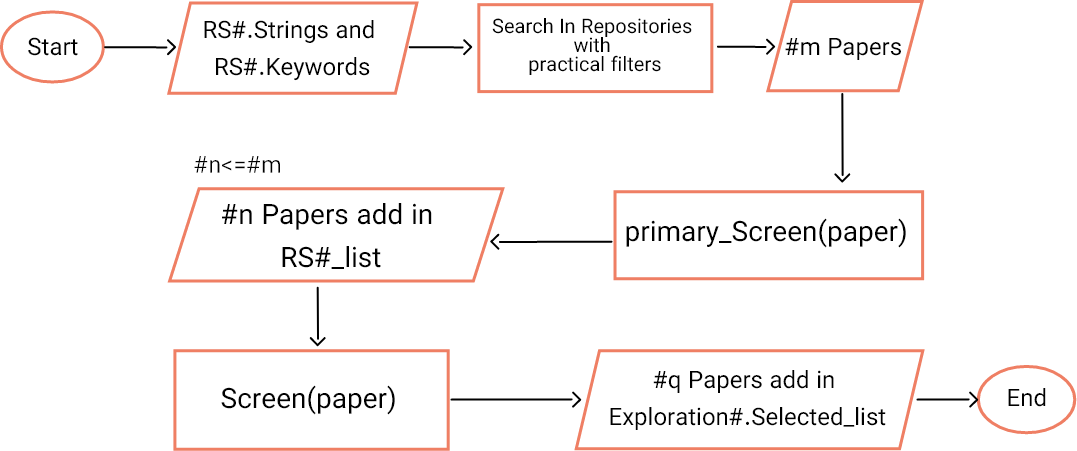}}	
	\caption{Search Algorithm}
\end{figure}

\subsection{Conducting the review}
\subsubsection{Domain of Studies}

\begin{figure}[t]
	\centering
	\subfigure[]{\includegraphics[width=0.54\textwidth]{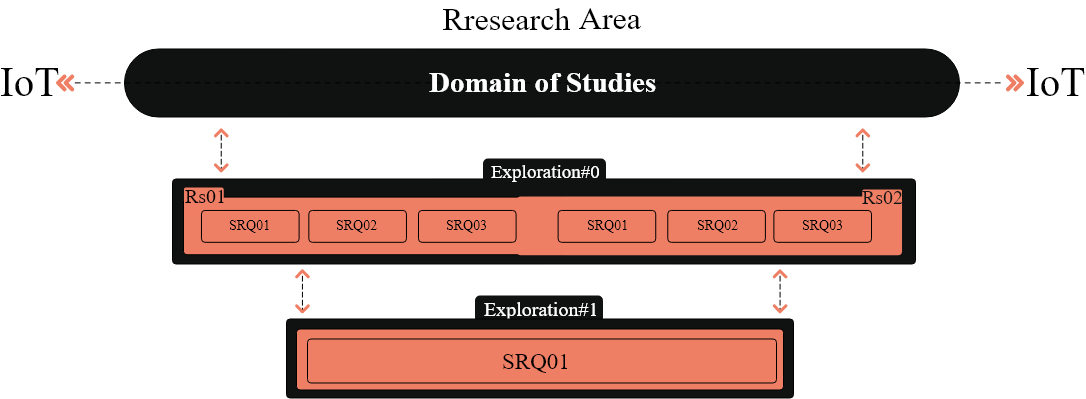}}
	\subfigure[]{\includegraphics[width=0.32\textwidth]{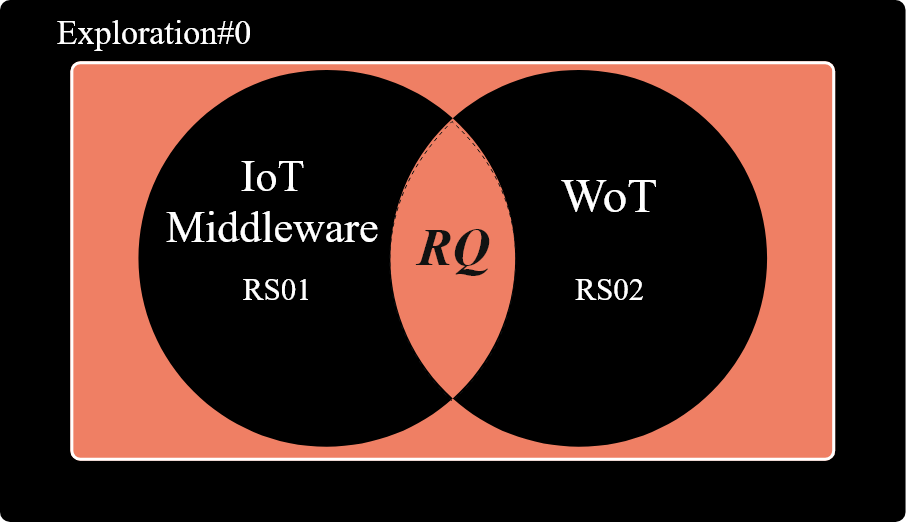}}	
	\caption{(a) Research conducting on protocol. (b) Common area of IoT-middleware and WoT}
\end{figure}

In an effective survey, a research scope must be identified and then focused on its restricted part. As depicted in figure 2.a, this paper indicates research scope as a DoS. Aforementioned research targets bounded DoS to certain topics, IoT fields and their relevant subjects. Research targets limit the DoS in two major fields as IoT-middleware and WoT. This survey investigates (possible) correlation between IoT-middleware and WoT. It is attempted to explore the DoS by defining some Exploration levels as follows.

\begin{itemize}
	\item \textbf{Exploration\#0:} The first level is defined to distinguish common area of WoT and IoT-middleware. This level involves two RSs to explore each field independently. RS01 is defined to discover information about IoT-middleware and their requirements. On the other hand, RS02 is defined to find out types of WoT architecture and its characteristics. The RQ of Exploration\#0 drives research process to identify some requirements for WoT architecture.
	\item \textbf{Exploration\#1:} This research level is responsible to find WoT platforms and assess them according to WoT requirements (figured out in prior Exploration level). RS01 is defined in this level to find the best WoT platforms or frameworks.
\end{itemize}

\subsubsection{Research Questions}

Table 1 indicates a regular categorization for entire RQs and SRQs in both Exploration levels. The main RQ of each Exploration level is stated as follows.    
\begin{itemize}
	\item \textbf{Ex\#0.RQ:}
	Exploration\#0 is proposed to focus on certain viewpoint of IoT studies that makes a correlation between two RSs and figures out WoT and IoT-middleware intersection (figure 2.b). The RQ of Exploration\#0 is defined to declare research trend to realize how many requirements of IoT-middleware are covered by characteristics of WoT architecture. To address this RQ, some SRQs are formed by means of two RSs.
	\item \textbf{Ex\#1.RQ:}
	The final purpose of Exploration\#1 is to discover WoT platforms among IoT platforms and compare them based on WoT requirements. The main RQ of this level is determined to evaluate WoT platforms from IoT-middleware perspective. The only SRQ of this level is defined to recognize some potential IoT-middleware platforms which adapted to WoT architecture.
\end{itemize}

\subsection{Reporting the Review}

\subsubsection{Primary Screen Results}

Search algorithm was executed on three popular repositories (i.e. Web of Science, Scopus, IEEE Xplore) for RSs in their own Exploration level and then found out the most relative primary studies. Figure 3.a shows the number of discovered papers per repositories. It is tried to exclude duplicate papers while executing primary\_Screen procedure, however, some selected papers were duplicate in RS selection list. These papers are detected and then eliminated from selection list.

As indicated in figure 3.b, final number of selected papers in every RS per Exploration level are obtained. It reveals that final number of selected primary studies for Exploration\#0 and Exploration\#1 is 156 and 184, respectively. Totally, this research reviewed and evaluated 340 papers.

\subsubsection{Screen Results}
By executing Screen procedure (as indicated in figure 1.b) on every RS individually, the most suited papers are selected to answer the research questions. Figure 3.c illustrates the final number of selected papers per Explorations in which, the Exploration\#0 contains 59 papers and the rest belong to Exploration\#1. In result, 123 papers are selected in this research that 12 of them are shared between both Explorations.

\subsubsection{Distribution Results}

Figure 3.d depicts the distribution of selected papers per publisher, reference type and year of publishing, respectively.
As indicated in this figure, IEEE has the most number of released papers for investigated fields in this research.
Reference type of the majority of selected papers are journal articles and conference papers as presented in figure 3.e. Two PhD and one master theses are included in the reviewed papers as well.
Figure 3.f reveals selected papers are mostly published in 2012 and 2016.

\begin{figure}[t]
	\centering
	\subfigure[]{\includegraphics[height=1.45in,width=2.55in]{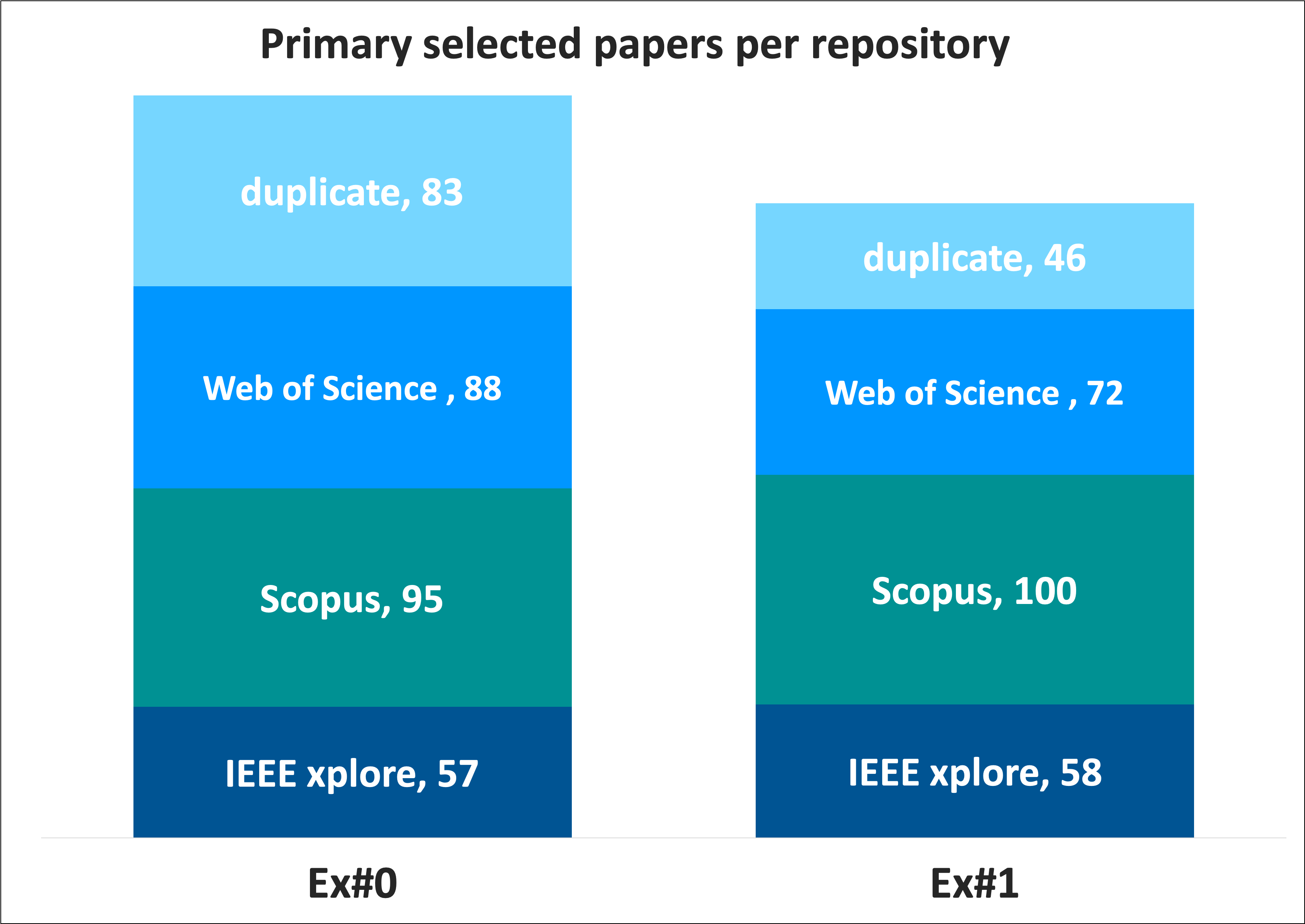}}
	\subfigure[]{\includegraphics[height=1.45in,width=2.55in]{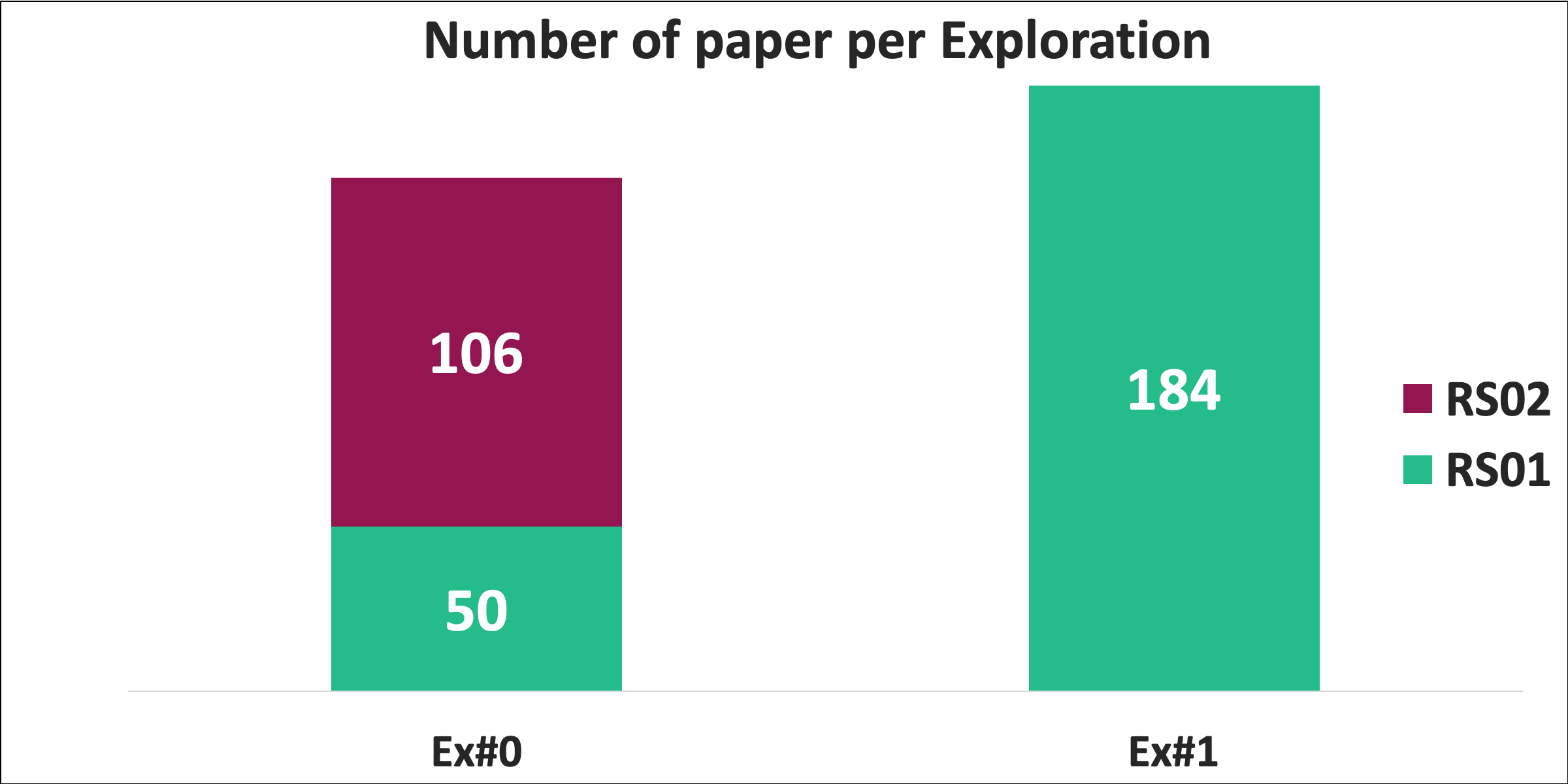}}
	\subfigure[]{\includegraphics[height=1.45in,width=2.55in]{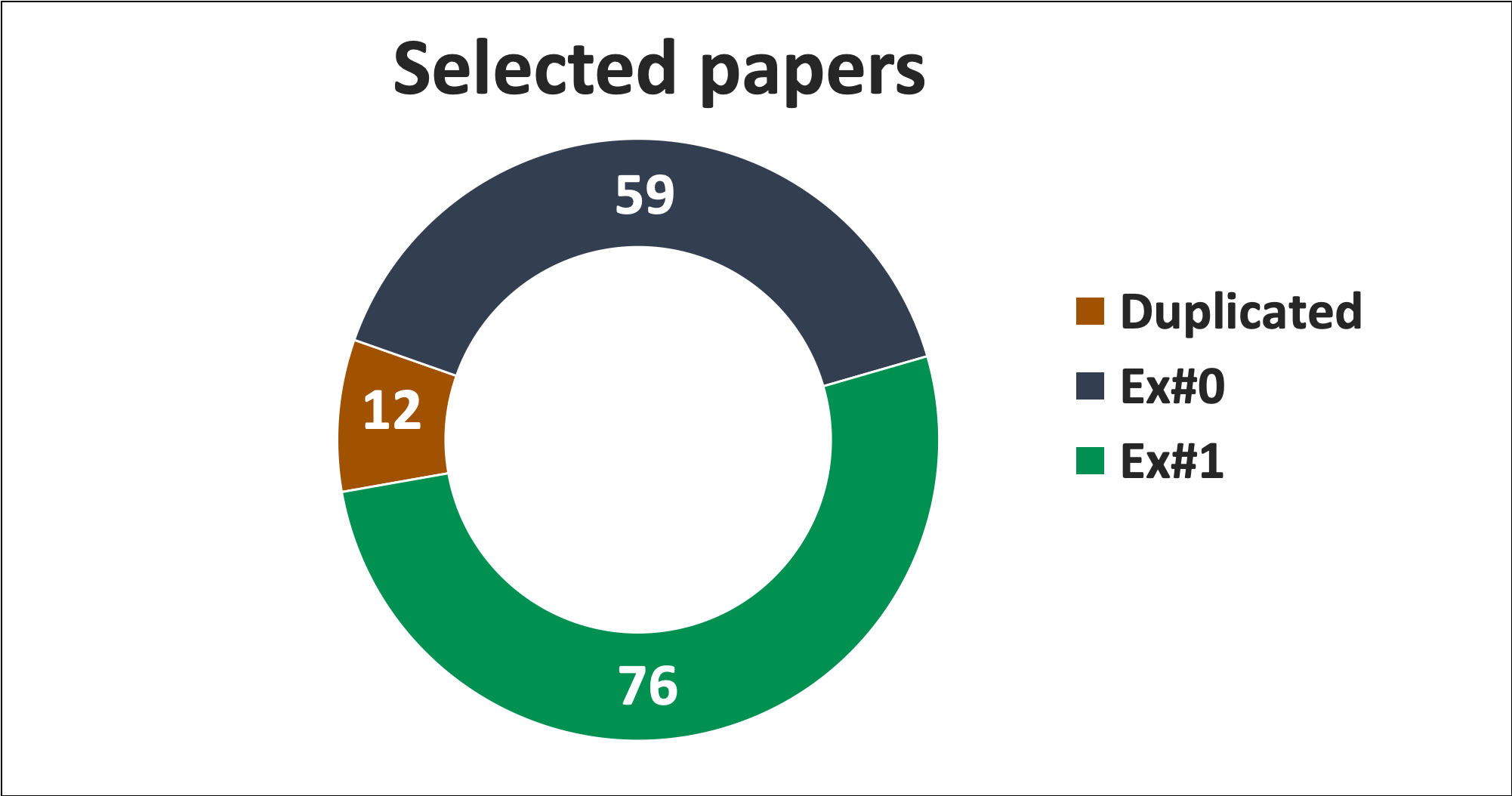}}
	\subfigure[]{\includegraphics[height=1.45in,width=2.55in]{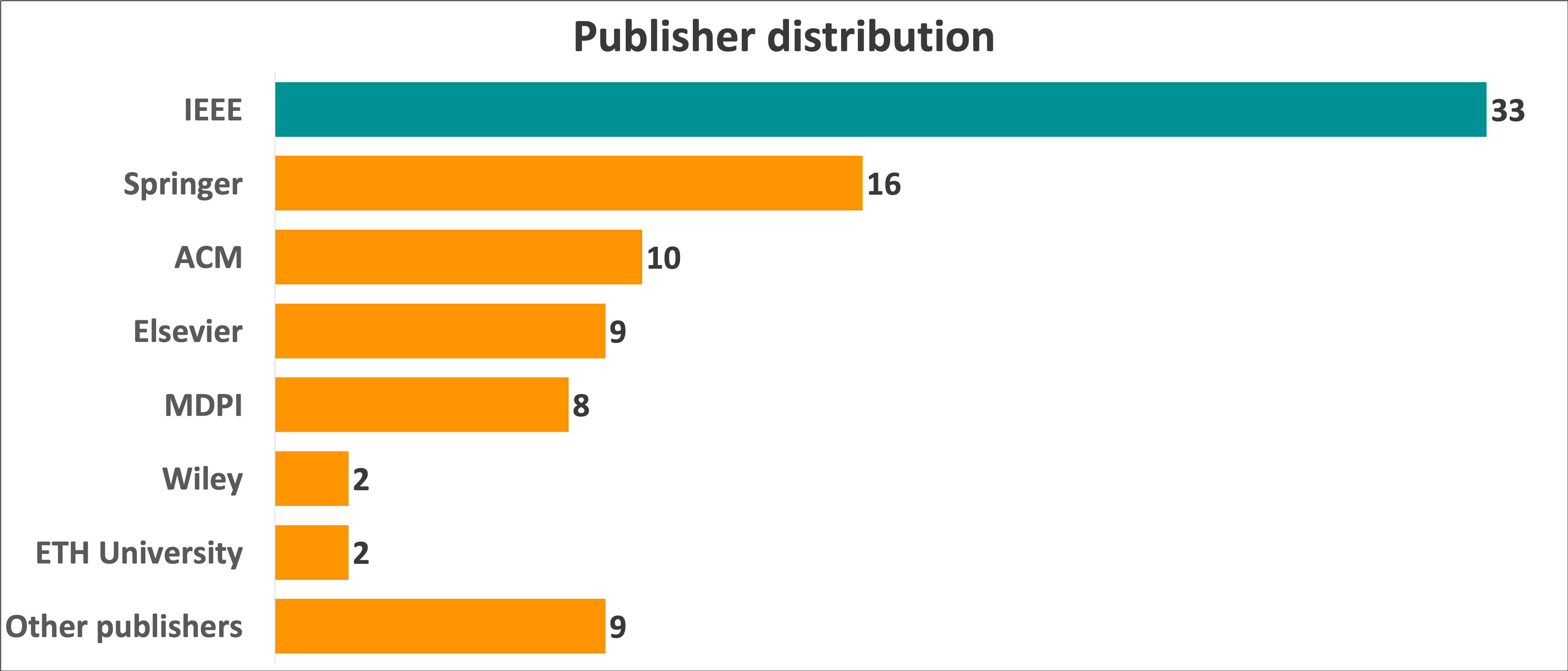}}
	\subfigure[]{\includegraphics[height=1.45in,width=2.55in]{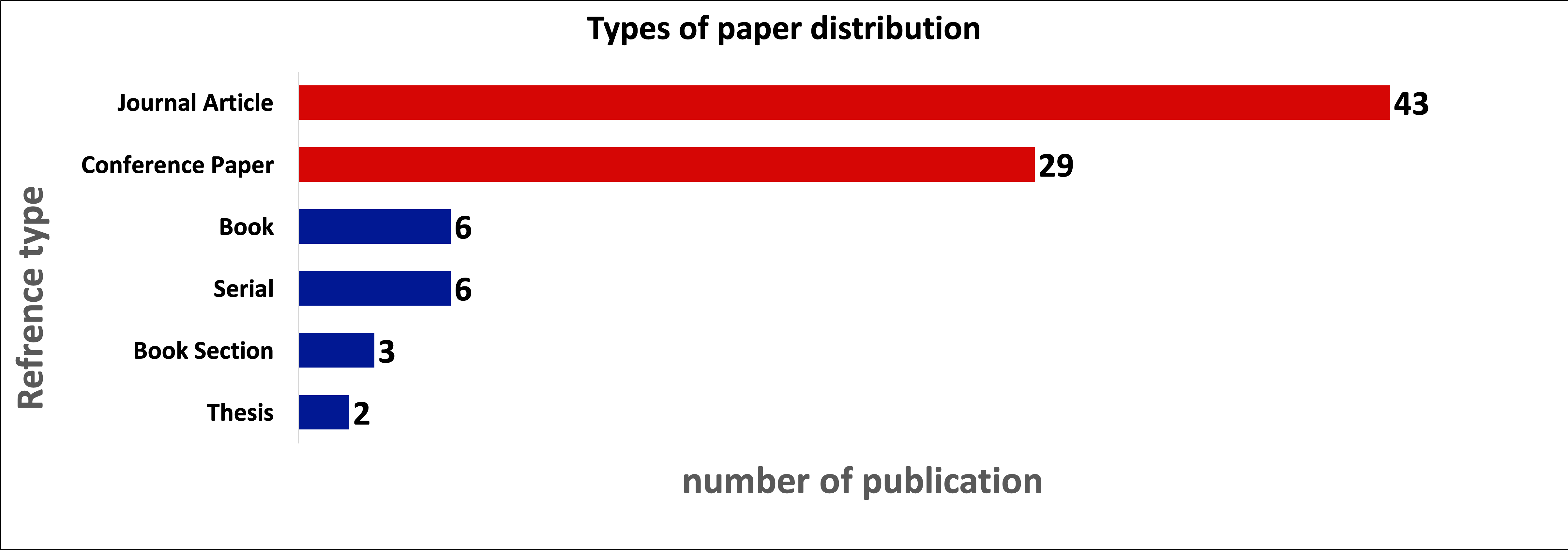}}
	\subfigure[]{\includegraphics[height=1.45in,width=2.55in]{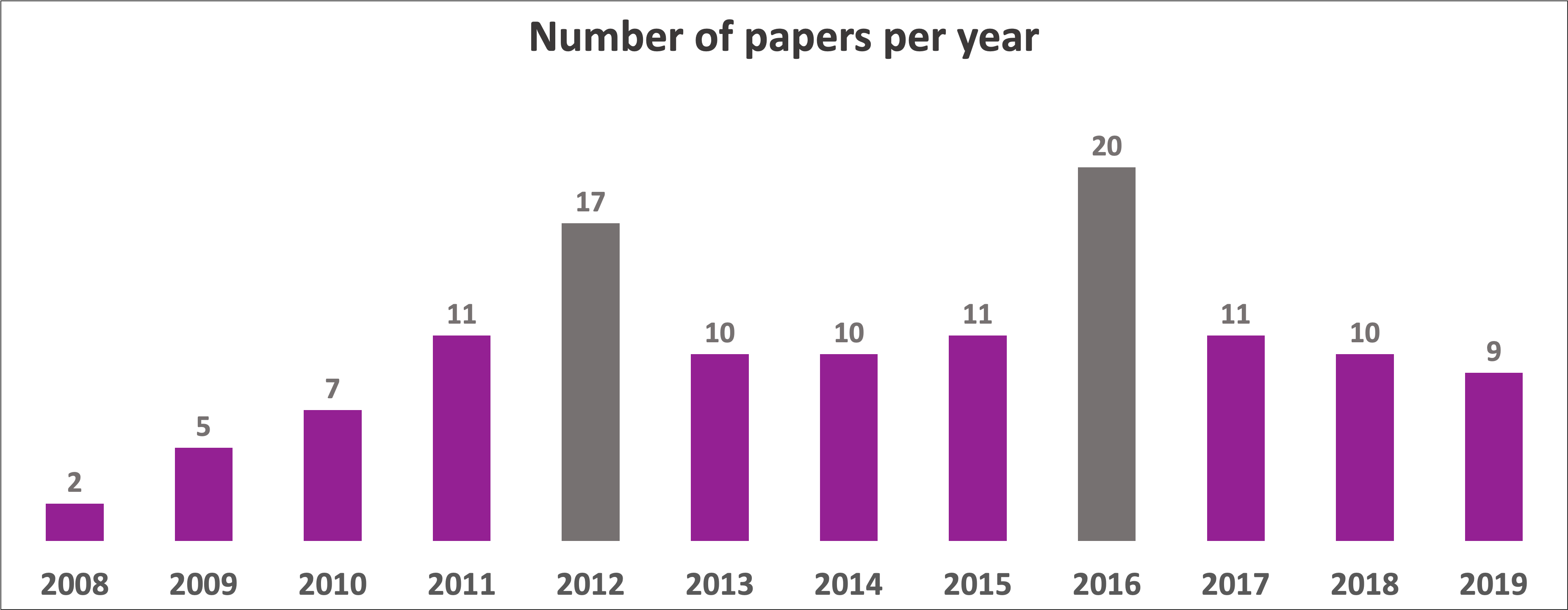}}	
	\caption{Research Results Diagrams}
\end{figure}

\section{IoT-Middleware Review}
\subsection{L0.RS1.SRQ01: What is the IoT-middleware?}

Nowadays, emerging IoT makes it feasible to use technology for building a smart environment under controlling and monitoring by humans or Artificial Intelligence (AI). Nevertheless, IoT could be defined as: “The worldwide network of interconnected objects uniquely addressable based on standard communication protocols” \cite{RN145}. In addition, The International Telecommunication Union (IERC) defined it as an enabler “allows people and things to be connected anytime, anyplace, with anything and anyone, ideally using any path/network, and any service” \cite{RN145}. As both aforementioned definitions realized, the main goal of the IoT is to enable effective communications and integrations between humans and environments in order to manage and control them.

Razzaque et. al. \cite{RN145} stated an industrial definition of IoT: "Industrial objects, or things, instrumented with sensors, automatically communicating over a network, without human-to-human or human-to-computer interaction, to exchange information and take intelligent decisions with the support of advanced analytics". Considering this definition, IoT not only enables communications between users and devices but it consists of communications between interconnected devices.

A basic architecture \cite{RN163} is presented to fulfill the main goal of IoT and disappear bound of digit and human realm. As indicated in figure 4.a, this architecture is formed in three layers: Application, Network and Perception. Obviously, Application and Perception layers are referred to human realm and digital devices, respectively. Thereby, the Network layer is responsible for making a seamless communication between two other layers. It could be possible to consider Network layer as a intermediate that hides complexities of communications between the interconnected devices and the users' applications. In other words, this middle layer plays a role of translator converting machine language to understandable human's language \cite{RN148,RN137}.

Consequently, a logical layer is necessary to abstract complexities of communications and aggregations among things or between users and things. This layer is called "Middleware". Middleware is a software layer located in the middle of Application and Perception (physical) layers in order to hide the details of different technologies from programmers and users \cite{RN131,RN145,RN148}. The primary goal of this software layer is to enable communication, and bring different (and incompatible) systems together \cite{RN137}. Middleware exempts programmers from developmental difficulties of each IoT component and facilitates developing applications on such distributed infrastructure like IoT. \cite{RN131,RN146,RN144}. In practice, Middleware provides common APIs (Application Programming Interfaces) to communicate various applications together \cite{RN145,RN146,RN137}.

\begin{figure}[H]
	\centering
	\subfigure[]{\includegraphics[height=2.5in,width=1.34in]{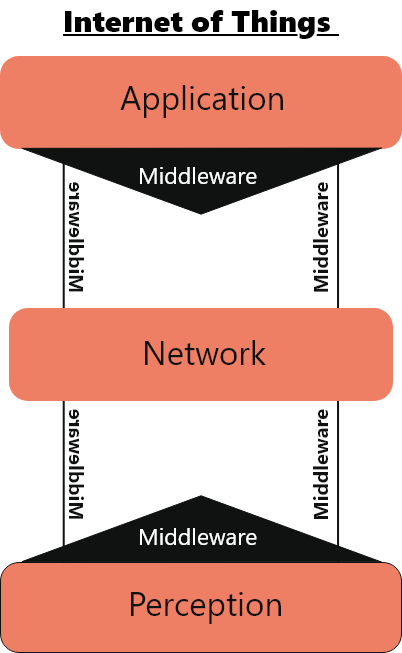}}
	\subfigure[]{\includegraphics[height=2.5in,width=2in]{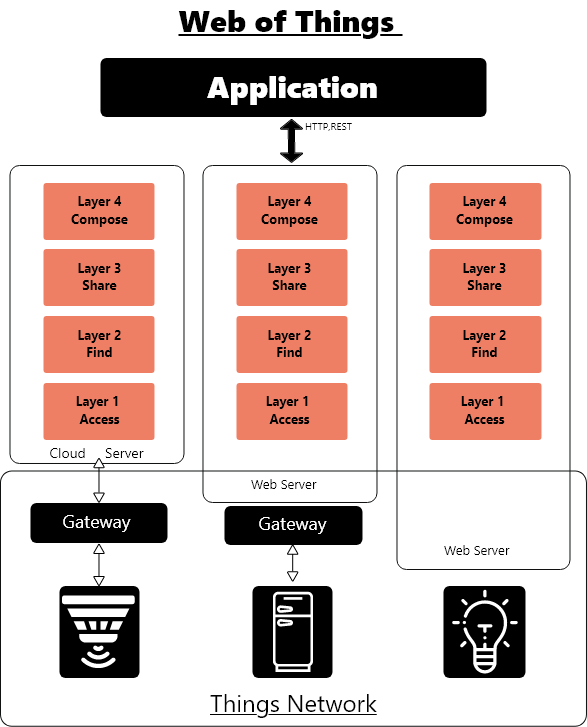}}
	\caption{(a) Basic IoT architecture. (b) WoT architecture}
\end{figure}

\subsection{L0.RS1.SRQ02: How many types of IoT-middleware exist?}

There are many types of IoT platforms that develop some applications to control and monitor smart devices. IoT-middleware is one of this type of platforms, which provides a common ecosystem to integrate and manage various user’s applications. In such platforms, applications are developed under common APIs that enable developers to focus on process of application development without knowing how sensors or actuators actually work. However, this type of platforms is produced with respect to certain scenarios \cite{RN137}.

IoT-middleware platforms can be investigated from variant perspectives such as architectural, technical characteristics, design approach, level of programming abstraction, and implementation domains \cite{RN138,RN148,RN145}. This paper classified architectures of IoT-middleware in two categories based on the level of abstraction:
\begin{inparaenum}
	\item Data abstraction 
	\item Device abstraction
\end{inparaenum}

\subsubsection{Data Abstraction Architecture}

In data abstraction architecture, the main task of the Middleware layer is major data processing such as acquisition, storage, and retrieve. In this type of architecture, data implies to corresponding functional units which should be pre-processed and prepared for next operations. Each functional unit is composition of (meaningful) data achieved from Perception (physical) layer. Data abstraction architecture includes sub types as follows.
\begin{description}
	\item [Event-based Middleware:] In event-based middleware, data transforms into events. firstly, events are generated by existent components in Perception or Application layer (event producers) and then they are propagated into their own interested subscribers (event receivers). Typically, event-based middleware acts as publish/subscribe model wherein event notifications are sent into subscribers in asynchronous manner \cite{RN148,RN145}.
	\item [Database-oriented Middleware:] Database-oriented middleware acts as a query engine to retrieve storing data on relational databases. In this middleware, IoT devices store their data on distributed databases. Thereby, developers must produce complex quires in SQL-like language and pass to middleware to retrieve and use required data in application development process \cite{RN145}.
	\item [Semantic-based Middleware:] In semantic-based middleware, data transforms to meaningful units by providing data descriptions. In fact, data is categorized into different types by adding some semantic values. This facilitates management of heterogeneous data by combining diverse data types together. Semantic descriptions also could simplify every process or interpretation of data in smart environment, for example every process of the objects discovery or knowledge extraction \cite{RN148}.  
	\item [Context-aware Middleware:] Context-aware middleware is a solution to process the immense and various data produced by IoT devices. Context-awareness includes two main tasks as context detection and context processing. Data is firstly collected and then the main factors  are identified in context detection. Context processing is responsible to extract useful data and process it. Finally, appropriate decision is made based on these processed results \cite{RN148}.
\end{description}
\subsubsection{Device Abstraction Architecture}

In device abstraction architecture, every existent IoT object in Perception layer (e.g. sensors, and actuators) abstracts as a reusable and understandable entity or service to be used in middleware. Abstraction in this level hides all of the implementation details like communication protocols or programming details on embedded devices. Nevertheless, functionality of every object is accessible through the common interfaces by middleware. Therefore, vendor developers exempt from considering the complexities of devised mechanisms in application development. Device abstraction comprises four sub types as fallows.

\begin{description}
	\item [Actor-based Middleware:] In actor-based middleware, various IoT objects are exposed as reusable actors distributed in entire network. This middleware consists of three basic layers as sensory layer, mobile access layer, and cloud layer. IoT objects in sensory layer abstract as actors and are then available via mobile access layer in the cloud. Each actor provides a service that could be available for other actors in case of need, for example storage service. The computation unit of the actor-based middleware is distributed on the network nodes. The key characteristic of this middleware is lightweight designing that enables it to be embedded in all layers \cite{RN138}.
	\item [VM-based Middleware:] The basic idea of VM-based middleware is modularization of programming environment in IoT architecture by virtualizing infrastructure objects. The applications separate in diverse modules distributed by middleware on network nodes. These nodes hold VMs (Virtual Machines) which are abstractions of the IoT objects. The main duty of the VM-based middleware is allocating each module into appropriate VM to interpret, execute, and finally rebuild results \cite{RN145}.
	\item [Agent-based Middleware:] In agent-based middleware, applications transform into separated modules in order to facilitate distribution of execution units on network through agents. Every part of Perception layer is abstracted as an agent and program modules run on entire network by agents. Every module of application can execute on agents so that it can retain its execution states while migrating between two agents. This approach simplifies decentralized systems issues by providing fault tolerance \cite{RN145}.  
	\item [Service-oriented Middleware:] Service-oriented middleware is based on SoA. Therefore, this type of middleware inherits significant characteristics of SoA like loose-coupling, service reusability, service composition and service discovery. In service-oriented middleware, software and application abstract as services presented through APIs. In fact, middleware hides entire software or hardware functionalities behind services. This feature candidates this type of middleware as the best approach for developers to develop IoT devices as services \cite{RN138,RN148,RN145}.
\end{description}

\subsection{L0.RS1.SRQ03: What are the requirements of the IoT-middleware?}

An IoT-middleware should support some key functionalities to ensure minimum necessities of the smart environments. These functionalities turn to requirements for IoT-middleware which should be satisfied. For example, security is a challenging issue that every type of smart environments (IoT-middleware) may be faced. Thus, IoT-middleware have to apply some security policies (e.g., authentication, encryption) to guarantee the security of the users’ data and devices. IoT-middleware requirements are categorized in two generic groups as: service requirements relevant to services offered by IoT-middleware and architectural requirements to approximate design-level to implementation-level \cite{RN148,RN145}.

\subsubsection{Service Requirements}
The service requirements are defined to ensure the performance of the system. These requirements are grouped in two sub-sets as functional and non-functional. The system functionalities and services are described in functional requirements, whereas non-functional requirements are utilized to assess Quality of Service (QoS). These requirements are stated as follows.

\begin{enumerate}[I.]
	\item \textit{\textbf{Functional Requirements}}
	\begin{description}
		\item [Resource Discovery:] Resource discovery plays a characteristic role in the IoT-middleware with large number of various heterogeneous components. Each device in the IoT-middleware which offers a resource must advertise itself to facilitate discovery process for other consumers or users. In practice, every device in a distributed system broadcasts its information to other neighbors. In this mechanism, IoT devices must be well-scaled \cite{RN159,RN145,RN137}.
		\item [Resource Management:] The IoT-middleware is responsible for managing resources to ensure they could handle users’ requests to provide acceptable responses. In other words, IoT-middleware must allocate resources fairly to avoid all types of probable contentions. In addition, every resource must be monitored to guarantee desirable resource utilization. This specifies the importance of the resource management on the IoT-middleware to enforce variant control services on the resources \cite{RN145,RN137}. 
		\item [Data Management:] Data management is a significant process in the IoT-middleware. In fact, a principal task of IoT-middleware is managing of data generated by IoT devices. This data, gathered by sensors/actuators or user applications, is often grouped in sets of identification data, descriptive data, and environmental data \cite{RN159}. Therefore, the IoT-middleware should provide data management services to take control of data flows including data acquisition, processing, and storage \cite{RN145}.
		\item [Event Management:] In the most cases of smart environments, there are some random and unpredictable events must be analyzed and pushed to subscribers spontaneously. The IoT-middleware should manage these events to ensure system dynamism. Moreover, the IoT-middleware is responsible to transform these simple observed events to meaningful ones and push them swiftly into their consumers \cite{RN148,RN145}.
		\item [Code Management:] Code management is ability of the IoT-middleware to facilitate code deployment through IoT environments. In practice, some services are needed to support code allocation and migration among IoT nodes. In code allocation service, a set of IoT nods is selected to allocate a particular application process. Code migration service is responsible for transferring code among IoT devices over the network \cite{RN145}. An IoT-middleware should also facilitate update operation while IoT devices are connected to the network \cite{RN137}.
	\end{description}
	\item \textit{\textbf{Non-functional Requirements}}	
	\begin{description}
		\item [Scalability:] An IoT-middleware essentially should support a network with large number of devices wherein they have various processing capabilities. It indicates the importance of fulfilling scalability requirement in such middlewares. By satisfying this requirement, IoT-middleware maintains its stability and service quality when the IoT’s network is growing exponentially \cite{RN148,RN145,RN137,RN163}.
		\item [Real-Time:] The efficiency of the IoT-middleware from end-users’ perspective approximately depends on real-time operation of services.  Real-time requirement is integral part of smart environments specially in elderly or patient care (healthcare) applications. To satisfy this requirement, middleware must reduce response time duration. The correctness of operation is not only important issue in real-time services. The processing time of such operation should also be considered in fulfilling this requirement \cite{RN148,RN145,RN137}.
		\item [Reliability:] Reliability of an IoT-middleware refers to its operational level even in case of failure. The ultimate goal of reliable middleware is to increase reliability of entire system including communication, data, technologies and smart devices. Leveraging load balancer and supporting variant communication methods could help to increase reliability level of middleware \cite{RN169,RN145,RN163}.
		\item [Availability:] The IoT-middleware services must be available in 24/7 basis specially in critical smart environments. It means a user could utilize middleware services anytime and anywhere. The availability requirement should accomplish in software and hardware levels. In the former level, services are simultaneously available for customers from any places. Latter one, refers to persistent availability of IoT devices. Availability and reliability should be considered as a necessity of a fault-tolerant system \cite{RN145,RN137,RN163}.
		\item [Security and Privacy:] In the last few years, the amount of data has been extremely growing by the emergence of IoT applications. Ensuring security of data always has been a challenge in the IoT-middleware. This huge amount of data on the typical IoT network attracts potential attacks such as Denial of Service (DoS), disclosure of sensitive data and etc. \cite{RN137}. To protect against these threats, a security policy is needed to enforce on all layers of IoT-middleware platform. This policy has to consider all the functional and nonfunctional blocks from Perception to Application layer \cite{RN145}. In addition, security policy should not only protect data against such types of attacks but also prevent possible data lost or destruction. Four security requirements including confidentiality, integrity, authentication, and non-repudiation need to be satisfied in order to guarantee secure IoT-middleware \cite{RN148}. 
	\end{description}
\end{enumerate}

\subsubsection{Architectural Requirements}

There is always a major gap between design and implementation level in development process of an IoT-middleware platform. Architectural requirements are defined to determine obscure points of middleware design for developers in order to facilitate implementation process. These requirements are described as follows.
\begin{description}
	\item [Extensibility:] The IoT is an innovating technology evolving gradually. This technology is faced many challenges and barriers that decrease its progress. An IoT-middleware architecture should be extensible to adapt itself with novel technologies to address such issues. For example, cloud/fog computing is one of the best technologies which could accommodate itself to IoT infrastructure and overcome the distributed computing challenges \cite{RN148,RN169}.
	\item [Modularity:] Designing and implementing of an IoT-middleware are often complex and difficult. Modularity always facilitates implementation process of the IoT-middleware and ensures the flexibility of middleware functions. However, middleware should manage entire modules simultaneously to maintain middleware consistency \cite{RN148}. 
	\item [Versatility:] In a smart environment, IoT infrastructure has two main characteristics as mobility of devices and dynamic networks \cite{RN145}. Smart objects in such environments have no fixed physical locations and/or network topologies. For example, whenever a device roams from one location to another, network topology is changing rapidly. Therefore, an IoT-middleware architecture has to adjust itself to new situation ceaselessly to handle such dynamism \cite{RN148,RN137}.
	\item [Multiplicity:] In IoT-middleware platforms, each service could be proposed by multiple devices or nodes with limited processing capability \cite{RN148}. Hence, IoT devices must communicate and collaborate to each other concurrently to present the most suitable services with respect to variant situations. The IoT-middleware should help users to select the best services to address their needs \cite{RN137}.
	\item [Intelligence:] Intelligent technologies must be leveraged to enable IoT devices or applications interact and communicate to each other without any external interventions. An IoT-middleware should use predictive and/or proactive strategy in architectural design to support intelligence requirement \cite{RN145}. 
	\item [Interoperability:] Interoperability is a crucial part of any IoT-middleware. Heterogeneous IoT components, in a both-side communication, must be able to realize their exchanged messages and services. These components should expose their functionalities through APIs to reduce developers’ efforts in making interoperable architecture \cite{RN137,RN148}. Some communication protocols must be defined to be compatible with IoT’s components. Interoperability is categorized into three classes as network, syntactic, and semantic. Network interoperability is responsible to define some network protocols to exchange information across network infrastructures. In syntactic interoperability, format and structure of encoding exchanged data is defined and semantic interoperability deals with meaning of exchanged data across middleware. In fact, some rules are defined in semantic interoperability for understanding meaning of information \cite{RN145,RN159}. An IoT-middleware must declare all above-mentioned interoperability classes to ensure the correctness of exchanging information via IoT components.
	\item [Abstraction:] Abstraction is the important level of architectural design. An IoT-Middleware system must be viewed from various levels of abstraction accurately (e.g. node level, system level, user level, and etc.). Developers have to design IoT system in different levels of abstraction to hide any module heterogeneities and communication complexities. Finally, developers should provide logical APIs to accelerate implementation of applications or services \cite{RN145,RN148}.
\end{description}
\section{Web of Things Review}
\subsection{L0.RS2.SRQ01: What is the Web of Things?}

IoT is not enough to meet all requirements of ubiquitous computing since the most issues in IoT are about Perception and Network layers. Although, IoT applications could be built on a network infrastructure but the main target of IoT is establishing connectivity in a vast range of smart environments \cite{RN63}. The first step towards the ubiquitous computing is connecting physical objects to the Internet and assigning unique addresses. The next step is to use the World Wide Web ecosystem to build applications for the IoT environments \cite{RN687,RN2}. 

Finally, the motivations towards IoT and a need to standard application platform opened the new research horizon to the WoT. In fact, WoT emerged as an common application platform to bridge the gap between IoT devices in digital realm and their applications in real world \cite{RN26}. In other word, “As the Web is to the Internet, so WoT is to the IoT” \cite{RN80}.

The aim of WoT is to utilize the current IoT tools and techniques to reuse, incorporate, and accommodate existing Web standards and technologies (e.g., HTTP, and REST) to build a seamless application platform to maximizes usage of the IoT \cite{RN90, RN123, RN687}.

D. Guinard defined WoT as “a refinement of the Internet of Things by integrating smart things not only into the Internet (network), but also into the Web Architecture (application)” \cite{RN2}. Its target is integrating the physical objects with Web technologies. These objects should be able to communicate with each other on IoT infrastructure by using Web protocols \cite{RN110}. In WoT, smart objects are considered as principal elements of the Web. Each physical object has a corresponding digital entity commonly referred to as “Web Thing” \cite{RN239, RN63}. Fundamentally, this level of abstraction hides the complexity and differences between various transport protocols used in IoT. It enables developers to focus on solving problems without any burden of how smart devices or transition protocols work \cite{RN687}. Figure 4.b depicts a WoT architecture developed by D. Guinard \cite{RN657}.

\subsection{L0.RS2.SRQ02: How many types of WoT architecture exist?}

The integration of smart things functionalities on distributed platforms is still the main issue in IoT applications development. Thus, WoT research area emerged as a promising solution to enable common platform to be used in IoT ecosystems. According to this solution, both IoT data and functionalities are abstracted as services which are accessible via uniform interfaces (i.e., APIs) \cite{RN124,RN102}. The primary task of these services is to manipulate Web resources by uniform set of operations \cite{RN63}. There are two major potential service-oriented approaches to correspond IoT devices to services: Service-oriented Architecture (SoA) and Resource-oriented Architecture (RoA).

\subsubsection{Service-oriented Architecture (SoA)}

SoA is designed to build flexible applications in loosely-coupled manner across distributed systems. Independent services interact to each other in a SoA-based application through a set of constraints \cite{RN42}, in which these interactions are happened by sending encapsulated messages. Technically, SoA is designed for large-scale systems such as enterprise organizations wherein clients and services communicate to each other in stateless fashion. This type of communication facilitates the implementation of scalable applications \cite{RN659}. SoA has an extensive array of additional standards (WS-* protocol stacks) which makes it as potential solution for scalable applications development e.g., WS-Discovery for discovering services, WS-Eventing for defining a publish/subscribe model on Web services, and WS-Security for guaranteeing the security of services.

SoA stands on four key enabling technologies comprising Simple Object Access Protocol (SOAP), Web Service Description Language (WSDL), Universal Description Discovery and Integration (UDDI), and Business Process Execution Language (BPEL) \cite{RN63}.

SOAP is an XML-based protocol defines the encapsulation standard for transmitting messages over the HTTP protocol. The basic element of SOAP messages is envelope which contains two XML sub-elements, called header and body. Header usually includes certain information about routing and QoS parameters and body involves message contents \cite{RN124,RN106,RN63}. SOAP message structure is described in a WSDL document, an XML document that describes syntactic of messages \cite{RN124}. In addition to Web services interfaces, the port type of endpoint (involving a set of operations) is defined in WSDL documents \cite{RN63}. These WSDL documents are registered in UDDI directory, an XML-based framework used for registration of Web services to facilitate service discovery process on World Wide Web \cite{RN124,RN63}. BPEL defines an interoperable composition model to extend service interactions into business transactions. In fact, BPEL composes various services to generate value-added business ones. BPEL can model service interactions in two ways of executable process and abstraction process. Contributors interactions management could be modeled in executable process. On the other side, the objective behaviors and process templates are described in abstraction process \cite{RN63}.

The basic idea of integrating SoA with real-world things is to make an ecosystem in which, smart devices abstract as services. In such ecosystem, well-accommodated services could be composed together to be utilized in special business use cases. OASIS has presented a lighter standard for such resource-constrained devices, called Device Profile for Web Services (DPWS) \cite{RNdp}. This standard is defined according to WS-* protocol stack. However, communications leverage HTTP protocol over UDP rather than TCP to decrease network traffic overheads and thereby, power consumption on resource-constrained devices \cite{RN131,RN124,RN102}.

\subsubsection{Resource-oriented Architecture (RoA)}

RoA is another solution to design and implement distributed applications. RoA as a type of RESTful architecture \cite{RN682} is a lightweight alternative architecture for SoA, in which fundamental units (every set of information models) abstract as resources \cite{RN124,RN42,RN659,RN106}.

There are four basic concepts defined in RoA: resources, Uniform Interface Identifiers (URIs), representation, and links. Every system component is defined as a resource identified by an unique address called URI \cite{RN687,RN63}. Each resource is described by representation items indicate data about states of resource. These resources finally connect together by links \cite{RN682}.

RoA has four key features, inherited from REST architectural style, as follows.

\begin{enumerate}
	\item \textbf{Addressability:} The information is exposed through some resources in RoA. Accessibility of these resources depends on system addressability, from user perspective. To simplify users’ access into the information, all the resources in RoA expose their information through URIs.
	\item \textbf{Uniform Interfaces:} The resources must be available to clients through uniform interfaces. There is a well-defined protocol, called HTTP, allows clients access to resources by an effective set of methods. These methods are created to reduce decoupling and optimize interactions between resources and clients. Some significant HTTP methods described as follows.
	\begin{description}
		\item [GET] is used to retrieve resource representation items.
		\item [PUT] is used to update or replace information of the resource representation items.
		\item [POST] is used to send some data to resources. 
		\item [DELETE] is used to remove representation of the resources.
	\end{description}
	\item \textbf{Stateless Interaction:} In RoA-based applications, occurred interactions among clients and servers, are stateless. It means, application states never be saved on the server-side, but they would be saved on the client-side in the form of information states (HTTP cookies). Technically, to serve a request, required information should be part of the request \cite{RN682}. 
	\item \textbf{Connectedness:} Connectedness is another term of “hypermedia as the engine of application state”, mentioned in Roy Fielding thesis \cite{RNN}, which means all the resources connect together by some links and forms. Clients could explore these links among resources to change their application states and interact with other services \cite{RN687,RN682,RN657}.
\end{enumerate}

The aforesaid features of RoA, make it potential candidate as basic architecture for WoT. However, implementation of HTTP protocol on resource-constrained devices is too expensive because of pull mode interactions. Therefore, another protocol is proposed to optimize resource consumption and to maximize efficiency on resource-constrained devices, called Constrained Application Protocol (CoAP). In practice, CoAP is implemented over the key features of RoA. CoAP uses the HTTP method to transfer its requests and responses. However, requests and responses are transmitted over the UDP instead of TCP. This causes reducing network overheads efficiently and thereby saving energy consumption on resource-constrained devices \cite{RN682,RN131, RNco}.

\subsection{L0.RS2.SRQ03: What are the specifications of WoT architecture?}
There are some abstracted characteristics in development process of WoT architecture. A IoT-middleware platform should support these characteristics in order to be considered as a WoT. These characteristics are proposed as follows.
\begin{description}
	\item [Resource discovery:]
	There are great number of smart objects, in form of resources, distributed through the WoT ecosystem. Users must query and discover their intended resources, either locally or globally, before starting interaction. For this reason, a search engine is necessary to search and discover such resources. Two approaches of developing a search engine for a WoT architecture are pull and push approaches. In the former, resources are discovered only upon receiving a user request and in the latter, spontaneous outputs information pushed to a search engine \cite{RN107,RN26,RN53}.
	\item [Integration things to the Web:] Two patterns exist to integrate smart objects as part of the Web.
	\begin{itemize}
		\item \textbf{\textit{Direct integration:}} In this pattern, smart objects firstly connect to Internet and take IP addresses. Then, devices would interact with the embedded Web server and finally abstract as resources. These resources are accessible directly via APIs and ultimately users could communicate with them through standard Web operations (e.g. GET, PUT, POST, and etc.).
		\item \textbf{\textit{Indirect integration:}} Some smart devices have very limited resources (e.g. power, memory, and etc.) to integrate directly into the Web. leveraging smart gateway as a medium addresses integrating this type of devices. Such gateway usually has sufficient resources to embed devices into the WoT. Cloud computing technology is an extended solution of indirect integration wherein, smart devices model as Web resources in the cloud and applications could access them through the Internet \cite{RN687,RN107,RN63}.
	\end{itemize}
	\item [Interoperability:] It is essential to smart devices utilize suitable Web standards for communications. Typical Web protocols, such as HTTP or WebSockets, may not be used on some resource-constrained devices  straightforwardly. Therefore, they should use more suitable protocols and standards like CoAP or MQTT \cite{RN26}.
	\item [Interaction API:] Every WoT resource must possess a certain API to provide a communication development point for WoT vendors. Appropriate APIs enable developers to create their custom and flexible applications or data visualizations without knowing complexity behind interoperability of devices \cite{RN26,RN126}.
	\item [Event and Alert:] In real-time scenarios, alerts must be sent simultaneously to the clients when a certain event occurred \cite{RN126,RN657}. For example, a fire detector sensor must send an alert message as soon as sensing a high temperature or unusual smoke.
	\item [Service semantic:] In WoT architecture, the description of services offered by physical things have to be documented. For example, WSDL is the default service description language used in the SoA \cite{RN53}.
	\item [Data semantic:] A semantic Web technology should be used for understanding meaning of exchanged data across the WoT architecture in order to facilitate parsing and integrating WoT data to other Web content \cite{RN53}.
	\item [Sharing:] A WoT architecture should permit users to share their smart objects with others and use the shared ones over the Web flexibly and securely. A main prerequisite to share smart objects is to use a set of common standards to encapsulate and transit data, in both side of devices and applications. To ensure security of sharing data, WoT should be capable to manage users' accessibility among different applications and devices \cite{RN53,RN687}.
	\item [Security and Privacy:] Sharing and exposing data are two security issues of any distributed architecture. The main issue of developing a WoT architecture is to ensure security of participants components and protect users’ privacy against misuse or malicious intervention. As a matter of fact, publishing sensitive information over the Web poses undesirable threats. Determining owner for each device and implementing authorization process in all levels of system access, are two practical approaches to ensure a secure WoT \cite{RN107}.
	\item [Physical mashups:] Web mashup is a desirable approach to boost WoT vendors to create complex applications by taking and combining information of several resources. Web mashup advantage for WoT architecture is to combine diverse physical resources and make a business service. Physical mashup is implemented in two ways as follows.
	\begin{itemize}
		\item \textbf{\textit{Physical-Virtual Mashup:}} In this type of mashup, applications are developed by combining information of both types of physical and virtual resources. A WoT application could be more effective when empowered with information of virtual and physical objects.
		\item \textbf{\textit{Physical-Physical Mashup:}} Participant services  contain physical resources in this type of mashup. Such mashup enables WoT applications with various functionalities of different smart devices in real world. It leads developers to focus on specification of application by using presented APIs \cite{RN107,RN106,RN687}.
	\end{itemize}
	\item [Controlling and Monitoring:] WoT platforms enable users to control and monitor environmental resources by collecting information from all existing sensors and actuators in the WoT platform \cite{RN126}.
\end{description}

\section{L0.RQ: How many requirements of the IoT-middleware are covered by characteristics of the Web of Things architecture?}
\subsection{Discussion}
WoT reuses Web technology to facilitate management and development of IoT infrastructures and/or applications. It is essential to accommodate Web standards to meet requirements of IoT in order to build a monolithic environment, involving physical and virtual components, established on the Web.

To fulfill requirements of IoT architecture by Web standards, elements of the Web architecture must be adapted. Web architecture leverages service-oriented model to provide a seamless environment of entities that offer their services through the Internet. WoT architecture would be built on both types of Web architecture i.e., SoA and RoA. SoA and RoA are accomplished in WoT architecture via SOAP-based and RESTful Web services, respectively. 

WoT architecture has to integrate and control enormous smart things. Thus, it must communicate amongst various components in loose-coupling manner to ensure scalability for heterogeneous smart devices. In addition, this type of architecture should be lightweight and flexible to guarantee tolerability on resource-constrained and embedded devices \cite{RN107}.

Table 4 indicates a comparison between two types of Web architectures, which are candidates for the WoT. This table compares both architectures considering several significant attributes (e.g. lightweight, flexibility, scalability and loose-coupling). The comparison results reveal RoA is more lightweight and flexible than SoA to build a WoT architecture \cite{RN659,RN107}.

There are several IoT-middleware solution to address IoT issues and challenges. Service-oriented middleware is one of the most effective and practical solution in which, many substantial issues are addressed and a vast range of IoT-middleware requirements are satisfied \cite{RN145}. As indicated above, the foundation of WoT architecture is based on the service-oriented middleware. Nonetheless, WoT is leveraging event-driven middleware, besides services-oriented, to support functional event management requirement.

By analyzing IoT-middleware requirements and comparing with characteristics of WoT architecture, it appears that WoT architecture is leveraging potentials of IoT-middleware to fulfill the requirements of smart environments. In this paper, some requirements are proposed for WoT platforms to evaluate them from IoT-middleware perspective. Table 3 indicates how IoT-middleware requirements map to WoT ones. These requirements are grouped in two basic sets as WoT-service requirements and WoT-architectural requirements.

\begin{table}[H]
	\centering
	\caption{How IoT-middleware Requirements map to WoT Requirements}
	\label{tab:my-table}
	\resizebox{0.6\textwidth}{!}{%
		\begin{tabular}{@{}cc@{}}
			\toprule
			\textbf{IoT-Middleware Requirements} & \textbf{WoT Requirements}                                                                               \\ \midrule
			\multicolumn{2}{c}{\textbf{Architectural}}         \\ \midrule
			Extensibility         & Integration                \\ \midrule
			Modularity            & Service paradigms          \\ \midrule
			\multirow{3}{*}{Versatility}         & \multirow{3}{*}{\begin{tabular}[c]{@{}c@{}}Service paradigms\\ Network\\  Integration\end{tabular}}                \\
			&                            \\
			&                            \\ \midrule
			\multirow{3}{*}{Multiplicity}        & \multirow{3}{*}{\begin{tabular}[c]{@{}c@{}}Resource discovery\\ Interaction API\\ Physical mashups\end{tabular}}  \\
			&                            \\
			&                            \\ 					\midrule
			Intelligence          & Data Semantic              \\ \midrule
			\multirow{4}{*}{Interoperability}    & \multirow{4}{*}{\begin{tabular}[c]{@{}c@{}}Network\\ Interoperability\\ Data format\\ Data semantic\end{tabular}} \\
			&                            \\
			&                            \\
			&                            \\						\midrule
			Abstraction                          & \begin{tabular}[c]{@{}c@{}}Service paradigms\\ Interaction API\end{tabular}                                       \\  \midrule
			\multicolumn{2}{c}{\textbf{Service-Functional}}    \\		\midrule
			Resource discovery    & Resource discovery         \\		\midrule
			\multirow{3}{*}{Resource management} & \multirow{3}{*}{\begin{tabular}[c]{@{}c@{}}Physical mashups\\ Controlling and Monitoring\\ Sharing\end{tabular}}  \\
			&                            \\
			&                            \\  \midrule
			Data management                      & \begin{tabular}[c]{@{}c@{}}Data format\\ Data Semantic\end{tabular}                                               \\  \midrule
			Event management      & Event and Alert management \\ 					\midrule
			Code Management       & Interaction API            \\					\midrule
			\multicolumn{2}{c}{\textbf{Service-Nonfunctional}} \\					\midrule
			Scalability           & Scalability                \\					\midrule
			Real-Time             & Real-Time                  \\					\midrule
			Security and Privacy  & Security and Privacy       \\					\midrule
			Availability          & Availability               \\					\midrule
			Reliability           & Reliability                \\ \bottomrule
		\end{tabular}%
	}
\end{table}

\subsection{Web of Things Requirements}
\subsubsection{WoT-Service Requirements}

WoT is grounded on the service-oriented middleware. Hence, WoT architecture must assure the requirements of this middleware.

WoT-service requirements are categorized into functional and nonfunctional subsets. Former requirements capture functions or services and latter ones capture QoS support (e.g., scalability, real-time).

\begin{enumerate}[i.]
	\item \textbf{\textit{Functional Requirements}}
	\begin{description}
		\item [Resource Discovery:] Smart environments integrate large-scale heterogeneous devices dynamically that interact in various networks (e.g., RFIDs, and WSNs). WoT architecture manages these devices as certain resources in order to provide atomic or composite services. These resources must be visible in network to communicate or compose. Therefore, a search engine must be considered to search and discover resources.
		\item [Physical Mashups:] In WoT architecture, resources must be composed and combined to provide seamless services, which enable developers to produce comprehensive Web applications. Importantly, these composite services provide common APIs to facilitate WoT applications development.
		\item [Controlling \& Monitoring:] A crucial task of WoT infrastructure is collecting data from sensors (e.g., value, status, location) or sending command to actuators (e.g.,switch on or off). Therefore, WoT application must control and monitor data flow between resources.
		\item [Event \& Alert Management:] A smart environment should be reactive to guarantee real-time QoS. In other words, it tends to react rapidly and attentively to special situation when a certain event is occurred. WoT architecture is responsible to manage and produce alerts and events as well as react accurately to special situation and notify user.
		\item [Sharing:] WoT applications users should access their things and share them over the Web. Therefore, WoT architecture should provide a secure mechanism to ensure security of data transformation. For this reason, it must encrypt messages and communications between things and users by use of security protocols. WoT architecture can use social networks (e.g., Facebook, Twitter) to authenticate owners of things and control their accesses to share information.
	\end{description}
	\item \textbf{\textit{Non-functional Requirements}}
	\begin{description}
		\item [Scalability:] WoT platforms should support the numerous heterogeneous devices. Scalability must be ensured as a QoS metric in order to maintain stability during the process of excessive requests and responses.
		\item [Real-Time:] In the world of information and data, real timing has a crucial role. A WoT application should be real-time to prepare users' responses in moment.
		\item [Security/Privacy:] A WoT architecture must ensure security of data and resources in entire layers of architecture. Privacy concerns on protecting of users' data from misuse and exposure.
		\item [Availability:] A WoT platform must be available all the times especially in critical environments, like smart cities or smart hospitals. Availability means, once a user needs to use a service or resource the system instantly receives a request, processes it real-time and replies a response.
		\item [Reliability:] A platform operation should remain and be stable during requests and responses processing. To achieve reliability, when the part of the system is down (as a component or service) the platform should be fault tolerant in order to remain operational. The reliability and availability features should work together to ensure platform fault tolerant.
		\item [Ease of deployment:] The simpler development of WoT platforms, the faster learning and promoting by developers. Therefore, the platform can be used in diverse environments with various requirements.
	\end{description}
\end{enumerate}
\begin{table}[t]
	\centering
	\caption{Comparison of SoA and RoA attributes (+ means low, ++ means medium, +++ means large)\cite{RN659}}
	\label{tab:my-table}
	\resizebox{0.3\textwidth}{!}{%
		\begin{tabular}{@{}lll@{}}
			\toprule
			\textbf{Attribute} & \textbf{SOA} & \textbf{ROA} \\ \midrule
			Lightweightness    & +            & ++           \\
			Flexibility        & ++           & +++          \\
			Scalability        & ++           & +++          \\
			Loose-coupling     & ++           & +++          \\
			Simplicity         & +            & +++          \\
			Standard           & ++           & +++          \\
			Tools available    & +++          & +            \\
			Security           & ++           & ++           \\
			Verbosity          & +++          & ++           \\
			Ambiguity          & ++           & ++           \\ \bottomrule
		\end{tabular}%
	}
\end{table}
\subsubsection{WoT-Architectural Requirements}	
To implement WoT architecture accurately, there are some requirements ensuring the performance of WoT platforms as follows.
\begin{description}
	\item [Service paradigms:] As mentioned before, WoT architecture is based on one of service-oriented model that implemented via RESTful or SOAP-based. Their consolidation, can be considered to take both advantages in WoT. To achieve it, there are two directions: 1) implementing SOAP-based via WS-* protocol stacks as central architecture and use the RESTful APIs for integrated things. 2) implementing RESTful architecture and merging into SOAP-based platform.
	\item [Network:] Firstly, smart devices should connect to Internet through a seamless network to send and receive information. Therefore, it needs to configure smart devices over a network protocol (e.g., RFID, WSN, Wi-Fi) and link them to Internet. 
	\item [Integration:] Smart objects should be integrated into the Web after allocating IP addresses through the network and connecting to Internet. There are two ways of integrating objects into the Web: integrating them into embedded Web servers directly or using a smart gateway as intermediate proxy that communicates to other components via installed drivers. on the other side, using cloud as an indirect integration allows Web platform to act as a gateway to integrate smart objects to WoT \cite{RN687}.
	\item [Interoperability:] Smart devices on the WoT need to use a Web communication protocol to transmit their information over the network. HTTP(s) is a well-known and efficient protocol to pull communication over the Web. However, its implementation on resource-constrained devices is not optimized. WoT architecture should present a lightweight protocol (e.g., CoAP) to disburden embedded resource-constrained devices. In addition, WoT architecture is capable to make push communication to enable alert messages. As such, WoT could use suitable protocols like WebSockets or MQTT.
	\item [Data Format:] There are various formats of exchanged data between components of WoT. JSON and XML are well-known and ordinary formats that used in the most of WoT platforms.
	\item [Data Semantic:] It refers to meaning of data transmitted over the Web. WoT architecture could use Semantic Web technologies (e.g., OWL, RDF) to understand meaning of data. 
	\item [Service Semantic:] WoT resources or services should be described in order to facilitate and accelerate discovery or composition process. Hence, WoT architecture provides a mechanism (e.g. WSDL or documented APIs) to store those descriptions.  
	\item [Interaction API:] To provide an abstraction layer to facilitate WoT applications development, WoT architecture should present APIs for integrated resources. In fact, platforms expose their services and resources functionalities through APIs to hide complexity of resources interoperability.  
	\item [Lightweightness:] WoT architecture must be lightweight to load on resource-constrained components. RESTful is more lightweight compare to SOAP-based architecture from implementation point of view.
\end{description}

\section{Research Analysis}
\subsection{L1.RS1.SRQ01: Which IoT-middleware platforms could be adapted to WoT architecture?}

IoT-middleware platforms provide interactive ecosystems using sensors, actuators, or any type of IoT objects. The main task of these platforms is to minimize the human interventions in management and control of diverse systems, based on their demands \cite{RN145}. As aforementioned, there are several types of IoT-middleware platforms designed to meet needs of different ecosystems. 

WoT platforms could be considered as special type of IoT-middleware platform leverage service-oriented middleware (RoA, SoA, or both) as a central core in order to integrate IoT objects to the Web. However, it could be possible to utilize multiple IoT-middleware types besides service-oriented such as event-based, or semantic-based. Some of these functional WoT platforms are assessed below.

\textit{SOCRADES} \cite{RN645} is a WoT platform leveraged a 5-layered architecture that designed to make IoT infrastructure accessible through service-oriented middleware. Practically, it abstracts the physical devices into embedded Web services by using DPWS. SOCRADES also supports the event-based middleware wherein events are produced by devices and published into Application layer by using WS-Notification standard. As a extension of this work, a discovery method presented in \cite{RN647} is proposed based on WS-Discovery to enable developers and business process designers to dynamically query, select, and use running instance of real-world services.

\textit{WoTKit} \cite{RN240,RN126} is a Java Web application that leverages the Spring framework, a popular development framework, for enterprise applications. WoTKit is designed with respect to basic requirements for lightweight toolkits: easy integration, visualization, processing ability and a RESTful interaction API.

\textit{SPITFIRE} \cite{RN277} is a semantic WoT framework that integrates sensors on Linked Open Data (LOD) cloud. This framework uses semantic Web technologies to provide descriptions for things and abstract them as semantic entities by using Resource Description Framework (RDF). SPARQL query is offered to search on top of such descriptions and a crawler scans periodically for semantic entities. The sensor nodes run 6LowPAN/IPv6 on the Network layer and communicate under CoAP transport protocol. They also use RESTful API to interact with outside world.

\textit{Paraimpu} \cite{RN354,RNp} is a social platform allows people to connect and share their information using virtual or physical things. Paraimpu enables users to discover and bookmark objects shared by other users/friends. This platform integrates the sensors and actuators using RESTful and communicate them through HTTP protocol.

\textit{GaaS} \cite{RN389} is a cloud-based platform, which integrates devices into service composition process to provide intelligent business process services. Its architecture maps three-layered cloud-computing stack into WoT architecture. Infrastructure as first layer of this architecture is responsible to provide RESTful APIs for smart things in order to abstract them as Web resources. In second layer, those resources map into business services to provide service composition and Business Process Management System (BPMS). Finally, end users can directly access the services on their own demands through UIs provided by the last layer.

\textit{ROSA} \cite{RN328} is a platform leverages the service model of Universal Plug and Play (UPnP) and RESTful architecture in order to provide a set of service management schemes for smart things management. This platform reuses the UPnP search protocol to provide a Web-friendly search mechanism and perform it on smart home applications.

\textit{WoT-SDN} \cite{RN332} is the WoT platform built on Software Defined Network (SDN). In this platform, resource subscription and mashup are developed through programmability feature provided by SDN. In addition, SDN could improve the security, efficiency and flexibility with complexity reduction. WoT-SDN is designed in three-layer. In first layer (infrastructure layer), WoT network is constructed by multi-domain SDN network and data flow is controlled by SDN controller. Resource coordinator and control layer is responsible to abstarct, store and coordinate the resources as a second layer. In third layer, applications could be developed by integrating of coordinated resources in previous layer.

\textit{HomeWeb} \cite{RN355,RN605} is a Web-based application framework for smart homes that uses RESTful architecture to alter home appliances as WoT resources. This framework supports concurrent accesses for multiple residents by providing request queue for each resource. Kamilaris et al. \cite{RN605} developed HomeWeb framework as an energy-aware smart home application that manages electricity waste actions through energy monitoring.

\textit{KNX-WoT} \cite{RN374} platform uses KNX gateway to integrate smart devices in WoT fashion. Although those devices connected to KNX network have no IP addresses, this platform develops a gateway to hide the complexity the KNX network and allows users to interact with KNX devices in WoT architecture.

\textit{WoVT} \cite{RN340} abstracts the physical entities (sensors and actuators) as virtual things. The architecture of Web of Virtual things (WoVT) comprises three layers of basic IoT architecture, which utilizes fog layer as an intermediate computing layer. WoVT servers locate in fog layer to manage communications between perception and cloud layers. Nonetheless, direct communications between end users and the perception layer is restricted to ensure privacy and security requirements.

\textit{EXIP} \cite{RN310} is a framework implemented over WoT architecture that uses Efficient XML Interchange (EXI) format for data exchange. EXI data format significantly reduces the size of XML messages when stored on disk or transferred over network. This framework presents design and implementation strategies for running an EXI processor on embedded devices. EXIP enables Web technologies to provide a WoT prototype implemented over CoAP and RESTful interfaces.

\textit{WTIF} \cite{RN325} is a two-layered framework based on the WoT architecture and Business Process Management (BPM). Device layer is in charge of abstracting smart things as Web resources by using a Web gateway. The next layer, BPM, is the core of business process modeling and execution management. WTIF is constructed via RESTful architecture leveraging SoA in which WS-BPEL is used as a service composer.

\textit{UbiQloud} \cite{RN606} is a cloud-based Platform as a Service (PaaS), provides Web-based UIs for developers which includes client APIs, smart gateway and social gateway. The smart gateway is responsible for communicating with integrated sensors and the social gateway consists of a set of APIs/widgets that can be used to integrate multi-service social login to client applications. UbiQloud is implemented in four modules (recognize, locate, connect, and share) in order to ease development of the platform by dividing the code into smaller pieces that could be developed separately.

\textit{WoT-ESN} \cite{RN313} is an energy-aware framework proposed over RESTful architecture in order to monitor and control home appliances. The utmost goal of WoT-ESN is to save energy of home appliances and thereby increase the monetary value.

\textit{ThingWorx} \cite{RNtw} is an industrial platform empowers organizations to transform how they leverage IoT solutions in their businesses. This platform contains a wide set of features, including multiple connectivity options, application development tools, analytics, and Augmented Reality (AR). All of these are built around the ThingModel which is a single real-time view of physical object. ThingModel is the key to the ThingWorx. It provides a consistent representation of "Things" to seamlessly tie all components of the platform into one.

\textit{µWoTO} \cite{RN256,RN414} is a healthcare platform, designed upon event-driven middleware that leverages WoT paradigm to create a smart space which advices people how to achieve a healthier lifestyle. µWoTO architecture is organized in three layers. In the first layer, platform deals with an ecosystem of sensors and actuators. WoT architecture, provides a set of functionalities to develop and deploy smart spaces in the second layer. In fact, this layer plays a crucial role to control and monitor data flow. It includes a consistent data flow from data acquisition by resources to output of events generations. The last layer, resource composition and orchestration, is used to enable µWoTO to offer platform services as well as additional services and functionalities in a RESTful fashion.

\textit{EVRYTHNG} \cite{RNev} is a cloud-based platform to store, share, and analyze data generated by physical objects. This platform abstracts objects (sensors, actuators, and NFC/RFID tags) as permanent digital identity namely Active Digital Identities (ADI), which allows authorized applications or users to access it via REST or MQTT APIs. ADIs provide a persistent and unique identity for each Web object to distinguish it from other objects. Thereby, physical objects in the form of ADIs could be part of Web mashup environment.
\begin{landscape}
	\begin{table}[p]
		\centering
		\caption{Summary of WoT platforms Assessments: Supported WoT Requirements, Part 1}
		\label{tab:platforms}
		\resizebox{1.4\textwidth}{!}{%
			\begin{tabular}{@{}lcccccc@{}}
				\toprule
				\multicolumn{1}{c}{\textit{\textbf{Platforms}}} &
				\multirow{2}{*}{\textbf{WoT-SDN}} &
				\multirow{2}{*}{\textbf{HomeWeb}} &
				\multirow{2}{*}{\textbf{KNX-WoT}} &
				\multirow{2}{*}{\textbf{WoVT}} &
				\multirow{2}{*}{\textbf{EXIP}} &
				\multirow{2}{*}{\textbf{WTIF}} \\ \cmidrule(r){1-1}
				\multicolumn{1}{c}{\textbf{Requirements}} &
				&
				&
				&
				&
				&
				\\ \cmidrule(l){1-7} 
				\multicolumn{7}{l}{\textbf{Architectural}} \\ \midrule
				Service paradigms &
				RESTful &
				RESTful &
				RESTful &
				RESTful &
				RESTful &
				RESTful \\
				Network &
				SDN &
				6LoWPAN, Zigbee &
				KNXnet/IP &
				Zigbee, BLE, Wi-Fi &
				Bluetooth &
				Zigbee, Bluetooth \\
				Integration &
				Indirect (Gateway) &
				Indirect (Gateway) &
				Indirect (Gateway) &
				Indirect (Cloud \& Gateway) &
				Direct &
				Indirect (Gateway) \\
				Interoperability &
				HTTP &
				HTTP &
				HTTP &
				HTTP(s) &
				CoAP &
				HTTP \\
				Data Format &
				JSON, XML &
				JSON,Text &
				JSON, XML &
				JSON &
				XML, XHTML, EXI &
				JSON, XML, ATOM \\
				Data Semantic &
				NS &
				NS &
				NS &
				NS &
				NS &
				NS \\
				Service Semantic &
				Resource Dictionary &
				WADL &
				API-Documentation &
				API-Documentation &
				API-Documentation (XHTML) &
				WSDL \\
				Interaction API &
				RESTful API &
				RESTful API &
				RESTful API &
				RESTful API &
				RESTful API &
				RESTful API \\
				Lightweight &
				No &
				Yes &
				Yes &
				Yes &
				Yes &
				Yes \\												\midrule
				\multicolumn{7}{l}{\textbf{Service-Functional}} \\ \midrule
				Resource discovery &
				NS &
				S (WS-Discovery) &
				S (DNSJava) &
				NS &
				NS &
				NS \\
				Physical Mashups &
				S &
				S (Mashup editor) &
				NS &
				NS &
				NS &
				S (WS-BPEL) \\
				Control \& Monitoring &
				S (Devices \& Services) &
				S (Devices) &
				S (Devices) &
				S (Devices) &
				S (Devices) &
				S (Devices \& Services) \\
				Event \& Alert Management &
				S &
				S (RMS) &
				S &
				S (Websockets) &
				S (CoAP observe) &
				NS \\
				Sharing &
				NS &
				NS &
				NS &
				NS &
				NS &
				NS \\  												\midrule
				\multicolumn{7}{l}{\textbf{Service-NonFunctional}} \\ \midrule
				Scalable &
				Yes &
				Yes &
				Yes &
				Yes &
				NI &
				NI \\
				Real-time &
				Yes &
				Yes &
				Yes &
				Yes &
				NI &
				NI \\
				Security \& Privacy &
				IA, Auth &
				NI &
				NI &
				C &
				NI &
				NI \\
				Availability &
				Yes &
				Yes &
				No &
				Yes &
				NI &
				NI \\
				Reliability &
				Yes &
				Yes &
				No &
				NI &
				NI &
				NI \\ \midrule
				\multicolumn{7}{l}{\textbf{Legend:} Supported (S), Not Supported (NS), No Information (NI), Confidentiality (C), Integrity (I), Availability (A), Authentication (Auth), Non-Repudiation (NR), Access Control (Ac)} \\ \bottomrule
			\end{tabular}%
		}
	\end{table}
	\begin{table}[p]
		\centering
		\caption{Summary of WoT platforms Assessments: Supported WoT Requirements, Part 2}
		\label{tab:platforms02}
		\resizebox{1.4\textwidth}{!}{%
			\begin{tabular}{@{}lcccccc@{}}
				\toprule
				\multicolumn{1}{c}{\textit{\textbf{Platforms}}} &
				\multirow{2}{*}{\textbf{SOCRADES}} &
				\multirow{2}{*}{\textbf{WoTKit}} &
				\multirow{2}{*}{\textbf{SPITFIRE}} &
				\multirow{2}{*}{\textbf{Paraimpu}} &
				\multirow{2}{*}{\textbf{GaaS}} &
				\multirow{2}{*}{\textbf{ROSA}} \\ \cmidrule(r){1-1}
				\multicolumn{1}{c}{\textbf{Requirements}} &
				&
				&
				&
				&
				&
				\\ \cmidrule(l){1-7} 
				\multicolumn{7}{l}{\textbf{Architectural}} \\ \midrule
				Service paradigms &
				SOAP-based &
				RESTful &
				RESTful &
				RESTful &
				SOAP-based (merge with RESTful) &
				RESTful \\
				Network &
				IEEE 802.15.4 &
				Bluetooth, ZigBee &
				6LoWPAN &
				Wi-Fi, Bluetooth &
				\begin{tabular}[c]{@{}c@{}}Zigbee, Z-wave,\\   Bluetooth\end{tabular} &
				Wi-Fi \\
				Integration &
				Direct (SunSPOT) &
				Indirect (Gateway) &
				Indirect (Cloud) &
				Direct &
				\begin{tabular}[c]{@{}c@{}}Indirect (Cloud \&\\   Gateway)\end{tabular} &
				Indirect (Gateway) \\
				Interoperability &
				DPWS &
				HTTP &
				CoAP &
				HTTP &
				HTTP &
				HTTP, HTTP-U, HTTP-MU \\
				Data Format &
				XHTML &
				JSON,HTML,CSV,KML &
				RDF &
				JSON, XML,Text &
				JSON, XML, ATOM &
				JSON, XML \\
				Data Semantic &
				NI &
				NI &
				RDF, OWL, SPARQL &
				NS &
				NS &
				NS \\
				Service Semantic &
				WSDL &
				Meta-Data &
				Meta-Data &
				API-Documentation &
				WSDL &
				\begin{tabular}[c]{@{}c@{}}API-Documentation\\   (XML)\end{tabular} \\
				Interaction API &
				RESTful API &
				RESTful API &
				RESTful API &
				RESTful API &
				\begin{tabular}[c]{@{}c@{}}RESTful API\\   (WS-Adaptor)\end{tabular} &
				RESTful API \\
				Lightweight &
				No &
				Yes &
				Yes &
				Yes &
				NI &
				Yes \\												\midrule
				\multicolumn{7}{l}{\textbf{Service-Functional}} \\	\midrule
				Resource discovery &
				S (WS-Discovery) &
				NS &
				S (SPARQL query) &
				NS &
				NS &
				S (UPnP) \\
				Physical Mashups &
				S (WS-BPEL) &
				S (Mashup pipes) &
				NS &
				S (Mashup composer) &
				S (WS-BPEL) &
				NS \\
				Control \& Monitoring &
				\begin{tabular}[c]{@{}c@{}}S (Devices \&\\   Services)\end{tabular} &
				S (Devices) &
				NS &
				S (Devices) &
				\begin{tabular}[c]{@{}c@{}}S (Devices \&\\   Services)\end{tabular} &
				\begin{tabular}[c]{@{}c@{}}S (Devices \&\\   Services)\end{tabular} \\
				Event \& Alert Management &
				S (WS-Brokered) &
				S (Apache Active MQ) &
				NS &
				NS &
				S (WS-Brokered) &
				S (GENA) \\
				Sharing &
				NS &
				S &
				NS &
				S &
				NS &
				NS \\													\midrule
				\multicolumn{7}{l}{\textbf{Service-NonFunctional}} \\	\midrule
				Scalable &
				Yes &
				Yes &
				Yes &
				Yes &
				Yes &
				Yes \\
				Real-time &
				Yes &
				Yes &
				NI &
				NI &
				Yes &
				Yes \\
				Security \& Privacy &
				CIA (WS-Security) &
				NI &
				NI &
				Auth &
				NI &
				CIA, NP,Auth \\
				Availability &
				Yes &
				NI &
				NI &
				NI &
				NI &
				Yes \\
				Reliability &
				NI &
				No &
				NI &
				NI &
				NI &
				NI \\		\midrule
				\multicolumn{7}{l}{\textbf{Legend:} Supported (S), Not Supported (NS), No Information (NI), Confidentiality (C), Integrity (I), Availability (A), Authentication (Auth), Non-Repudiation (NR), Access Control (Ac)} \\ \bottomrule
			\end{tabular}%
		}
	\end{table}
	\begin{table}[p]
		\centering
		\caption{Summary of WoT platforms Assessments: Supported WoT Requirements, Part 3}
		\label{tab:platforms03}
		\resizebox{1.4\textwidth}{!}{%
			\begin{tabular}{@{}lcccccc@{}}
				\toprule
				\multicolumn{1}{c}{\textit{\textbf{Platforms}}} &
				\multirow{2}{*}{\textbf{UbiQloud}} &
				\multirow{2}{*}{\textbf{WoT-ESN}} &
				\multirow{2}{*}{\textbf{ThingWorx}} &
				\multirow{2}{*}{\textbf{µWoTO}} &
				\multirow{2}{*}{\textbf{EVRYTHNG}} &
				\multirow{2}{*}{\textbf{}} \\ \cmidrule(r){1-1}
				\multicolumn{1}{c}{\textbf{Requirements}} &               &             &                        &                 &                        &  \\ \cmidrule(l){1-7} 
				\multicolumn{7}{l}{\textbf{Architectural}}     \\ \midrule
				Service paradigms                         & RESTful          & RESTful        & RESTful                   & RESTful            & RESTful                   &  \\
				Network &
				\begin{tabular}[c]{@{}c@{}}Feig LRU, RealSens\\   (WSNs)\end{tabular} &
				Zigbee, Bluetooth &
				Wi-Fi, GSM &
				Bluetooth, Wi-Fi, BAN &
				\begin{tabular}[c]{@{}c@{}}Zigbee, Bluetooth,\\   NFC, Wi-Fi\end{tabular} &
				\\
				Integration &
				\begin{tabular}[c]{@{}c@{}}Indirect (Cloud \&\\   Gateway)\end{tabular} &
				Indirect (Gateway) &
				Indirect (Cloud) &
				Indirect (Gateway) &
				Indirect (Cloud) &
				\\
				Interoperability                          & HTTP, XMPP    & HTTP        & CoAP, MQTT, HTTP, XMPP & HTTP(s)         & CoAP, MQTT(s), HTTP(s) &  \\
				Data Format                               & JSON, XML     & NI          & JSON,XML,HTML,CSV,Text & JSON, XML (CAP) & JSON                   &  \\
				Data Semantic                             & NS            & NS          & NS                     & NS              & Semantic data store    &  \\
				Service Semantic &
				\begin{tabular}[c]{@{}c@{}}API-Documentation (XML\\   or JSON)\end{tabular} &
				API-Documentation &
				API-Documentation &
				\begin{tabular}[c]{@{}c@{}}Service Register (OSGI\\   bundle)\end{tabular} &
				SDK, API-Documentation &
				\\
				Interaction API                           & RESTful API   & RESTful API & RESTful API            & RESTful API     & RESTful API            &  \\
				Lightweight                               & NI            & No          & Yes                    & Yes             & Yes                    &  \\  \midrule
				\multicolumn{7}{l}{\textbf{Service-Functional}}    \\ \midrule
				Resource discovery                        & NS            & S           & S (SQUEAL)             & S (UPnP)        & S                      &  \\
				Physical Mashups                          & NS            & NS          & S (Mashup builder)     & S               & S                      &  \\
				Control \& Monitoring &
				S (Devices) &
				S (Devices) &
				\begin{tabular}[c]{@{}c@{}}S (Devices \&\\   Services)\end{tabular} &
				S (Devices) &
				\begin{tabular}[c]{@{}c@{}}S (Devices \&\\   Services)\end{tabular} &
				\\
				Event \& Alert Management &
				\begin{tabular}[c]{@{}c@{}}S (XMPP over\\   Websockets)\end{tabular} &
				NS &
				S (Websockets) &
				S (Webhook) &
				S (Websockets) &
				\\
				Sharing                                   & S ( O-Auth)   & NS          & S (O-Auth)             & NS              & S (O-Auth 2)           &  \\		\midrule
				\multicolumn{7}{l}{\textbf{Service-NonFunctional}}        \\ \midrule
				Scalable                                  & Yes           & No          & Yes                    & Yes             & Yes                    &  \\
				Real-time                                 & Yes           & No          & Yes                    & Yes             & Yes                    &  \\
				Security \& Privacy                       & CIA, Auth, NR & NI          & CIA, Auth, Ac          & CI, Auth        & CIA, Auth, NR, Ac      &  \\
				Availability                              & Yes           & Yes         & Yes                    & NI              & Yes                    &  \\
				Reliability                               & NI            & Yes         & Yes                    & Yes             & Yes                    &  \\ 	\midrule
				\multicolumn{7}{l}{\textbf{Legend:} Supported (S), Not Supported (NS), No Information (NI), Confidentiality (C), Integrity (I), Availability (A), Authentication (Auth), Non-Repudiation (NR), Access Control (Ac)} \\ \bottomrule
			\end{tabular}%
		}
	\end{table}
\end{landscape}
\subsection{L1.RQ: How many WoT requirements are supported by WoT platforms?}
\subsubsection{Supported Requirements}
Strength of WoT architecture is based on satisfying substantial requirements of IoT ecosystems. Subsequently, the following requirements would be evaluated with respect to aforementioned WoT platform in table 5-7 to apprise WoT architecture.
\begin{enumerate}[I.]
	\item \textbf{\textit{Service-Functional}}
	\begin{description}
		\item [Resource discovery:] WS-Discovery, UPnP, and SPARQL query are the most popular discovery mechanisms utilize in WoT platforms (6 of 14  supported platforms). WS-Discovery is often used by SOAP-based platforms. ROSA platform \cite{RN328} proposed an optimized discovery mechanism based on the UPnP whereas KNX-WoT platform \cite{RN374} used DNSJava to perform resource discovery. Other platforms usually did not propose an optimized mechanism.
		\item [Physical Mashups:]  There are some tools such as mashup editor, mashup pipes, and WS-BPEL to combine several physical resources and build a composite service. Some platforms (10 of 17) supported this feature to build composite services and develop their applications.
		\item [Controlling \& Monitoring:] Almost entire WoT platforms (16 of 17) supported this requirement as integral part of their architectures. Some platforms (7 of 16) control and monitor data flows in device and service levels and rest of them support this feature in only device level.
		\item [Event \& Alert management:] WoT platforms have to use some protocols such as MQTT or XMPP to emit real-time events into users. These types of protocol make a both side interactive communications between users and servers. There are some popular technologies (e.g., WS-Broker and WebSockets) that provide APIs in order to enable WoT platforms support this type of communications. Majority of WoT platforms (13 of 16 supported platforms) use these technologies to fulfill this requirement.
	\end{description}
	\item \textbf{\textit{Service-Nonfunctional}}
	\begin{description}
		\item [Scalability:] A WoT platform should capable to support large-scale heterogeneous devices. Most of the platforms (14 of 17) evaluated WoT architecture in scalable environment and estimated its performance under controlled circumstances. Their results were desirable in some cases. Practically, WoT architecture could be used in scalable environments to handle physical resources as logical ones.
		\item[ Real-time:] This requirement is supported in most case of WoT platforms (12 of 17). These platforms proposed an evaluation to indicate the time between sending requests and receiving responses. The results were under 900 milliseconds at worst. Although, sampling and testing were experimented under controlled conditions but WoT platforms have potential to provide real-time communications in pragmatic conditions.
		\item [Security and Privacy:] Providing security and privacy of users’ information is an integral part of any WoT platforms. Some platforms (9 of 17) proposed a security mechanism to protect users’ information against any threats and malicious interventions. These mechanisms, meet entire or partial essential security requirements such as confidentiality, integrity, availability, authentication, non-repudiation, and access control.
		\item [Availability:] Most of the platforms did not report any details to evaluate availability of their services. However, some of them (9 of 17) presented a chart that reveals performance of services in resource utilization peak. In addition, some platforms proposed an efficient mechanism to maintain service operations and protect platforms against DDoS attacks.
	\end{description}
	\item \textbf{\textit{Architectural}}
	\begin{description}
		\item [Service paradigms:]Majority of WoT platforms (15 of 17) choose RESTful for service paradigm. However, SOAP-based as another service paradigm option is used more in enterprise systems. In practice, implementing technology of RESTful is simpler and more lightweight than SOAP-based service paradigm. It could be possible to merge two paradigms in order to leverage both robustness in WoT enterprise projects. For example, GaaS platforms is developed on two service paradigm models. 
		\item [Network:] WoT platforms connect Web Things together through different network technologies. Bluetooth, Zigbee, 6LowPAN are the most common wireless networks used in examined platforms. The common key characteristic of these networks is low-power consumption which makes them potential candidates to interconnect resource-constrained objects. SDN is an interesting new technology, interconnected Web Things in WoT-SDN platform \cite{RN332}. Body Area Network (BAN) is a certain network specialized for sensors used in healthcare systems such as µWoTO platform \cite{RN256}.
		\item [Integration:] Platforms mostly use indirect patterns (14 of 17) to integrate things to Web. These platforms ordinarily use gateway (8 of 14) for environments by a limited number of objects such as smart home. Cloud as another type of indirect pattern, would be used in large-scale ecosystems. Some platforms (3 of 14) leverage both gateway and cloud solutions as integration pattern in order to support vast range of devices. 
		\item [Interoperability:] HTTP as a well-known protocol is the most commonly used in WoT platforms (14 of 17) that makes communications among Web resources by transmitting messages over TCP connections. Although this protocol is an effective one, but it is barely optimized for resources-constrained devices. CoAP as the best alternative, transmits messages by using same HTTP methods over the UDP connection. Some platforms used MQTT and XMPP beside HTTP or CoAP to handle event-driven middleware.
		\item [Data format:] Both JSON and XML are considered as common data types which format Web massages. Generally, data in JSON format is more human-readable than XML. However, XML supports more data types (e.g., image, charts, graphs) than JSON. Hence, these data formats are frequently used in WoT platforms. CSV is another data type that used by platforms to format their messages. KML is used in WoT platforms (e.g. WoTKit) to transmit and visualize geographical data. 
		\item [Service semantic:] WoT platforms (9 of 17) store information of services as documents in regular formats such as XHTML, XML, or JSON. However, other platforms preferred to store this information as meta-data. In such platforms built on SOAP-based paradigm, service information is stored in WSDL documents.
		\item [Interaction API:] Entire WoT platforms (17 of 17) use RESTful API in abstraction layer to facilitate development process for vendors. SOAP-based platform use WS-Adaptor to convert their APIs into RESTful ones.
		\item [Lightweightness:] Technically, RESTful architectures are more lightweight than SOAP-based ones. Most of the RESTful platforms (12 of 17) are generally lightweight to consume resources such as CPU, memory, storage. In practice, implementation technology (programming language, framework, executing environments, or etc.) has a key role to determine lightweightness of the platforms.
	\end{description} 
\end{enumerate}

\subsubsection{Not-Supported Requirements and Open issues}

Development process of WoT requirements is an endless cycle during the lifetime of this technology. In spite of existing efforts, some of aforementioned requirements could be still considered as open challenges in WoT architecture. Addressing these issues can be eventuated to optimize WoT platforms. The most significant of them are stated as follows.

\begin{description}
	
	\item [Data semantic:] One of the important research areas in the Web and WoT is to make meaningful data and transmit it over the Web. Existing technologies (OWL, RDF) in data semantic are designed for just semantic Web resources. Other efforts in semantic platforms are unable to build a coherent mechanism to turn Web Things into semantic ones. As indicated in Table 6, only one platform (SPITFIRE \cite{RN277}) just focused to bring semantic into WoT.   
	\item [Resource discovery:] There is no optimized discovery mechanism for WoT resources. Most of the existing discovery mechanism (e.g. WS-Discovery, UPnP, SPARQL) are not efficient to explore Web Things in large-scale environments. WoT architecture should manage and control gigantic number of Web Things in such environments flawlessly to be considered as future of distributed systems. This is not feasible unless a efficient resource discovery mechanism will be developed. Therefore, fulfilling this requirement is still an open issue as future work.
	\item [Physical Mashups:] Development of this requirement has a key role to deploy WoT technology. Physical mashups accelerates vendors to build and deploy WoT applications in smart environments. It needs to consider some elements such as QoS in composition of several resources to make a logical service. There are variant mashups algorithms which are interesting to research and implement in WoT architecture \cite{RN174}.
	\item [Sharing:] Sharing users’ information to third-party applications in secure manner is always a challenging issue in WoT platforms. There are some standards that grant WoT applications to access users’ information such as OAuth. In WoT platforms, users could share their Web Things to another third-party Web services. A standard is required to guarantee secure accessibility of information and devices between users and applications.
	\item [Security and Privacy:] Ensuring security and privacy in IoT and WoT architectures have been always serious concern. Providing a comprehensive security mechanism to protect devices and information against malicious attacks is still an open issue. For example, authentication process would be progressed and optimized in WoT architecture. Moreover, access control as a security requirement could be altered in order to accommodate to WoT users and devices. To guarantee confidentiality and integrity of WoT information, data encryption algorithms should be optimized efficiently to execute on resource-constrained devices.
	\item [Availability:] WoT platforms may face to some issues in providing available services or resources in critical environments. To implement available platforms, it must be assured about operational precision in entire components of architecture. Hence, this requirement could be considered as a significant research topic.
	\item [Reliability:] This requirement has a vital role in critical environments and applications such as healthcare, medical care, and elderly care. In spite of wireless networks and communications progress, reliability is still a main issue in such systems. There are many aspects have to be considered to design and build a reliable platform. Providing an accurate and secure message transmutation within such platforms is only one aspect of satisfying this requirement.  Therefore, the importance of this issue is highlighted in implementation and investigation.
\end{description}

\section{Conclusion}

Reusing Web technology to control remotely cyber-physical systems is an impressive solution. However, its efficiency is still questionable from IoT-middleware perceptive to support essential requirements of IoT ecosystems. Nevertheless, there are not exact touchstones for WoT architecture to evaluate these requirements. This paper tries to analyze IoT-middleware and WoT discipline in order to map IoT-middleware requirements to WoT requirements by leveraging SLR methodology. This systematic review evaluated 340 papers totally and selected 123 potential studies to address main questions of this research. To accomplish this review, 17 potential WoT platforms (academic and industrial initiatives) are selected and compared against WoT requirements to accentuate strengths and weaknesses of WoT architecture to satisfy IoT ecosystems requirements.

Although WoT architecture satisfies almost entirely architectural and service-functional requirements but also it needs serious efforts to satisfy service-nonfunctional requirements e.g., security and privacy. Furthermore, it is strongly recommended that some WoT service-functional requirements such as resource discovery and physical mashups would be further investigated and developed. WoT architecture could benefit from other types of IoT-middleware besides service-oriented and event-driven to meet more requirements. Future works would investigate WoT foundation levels to concrete a standard infrastructure with capability of developing and deploying various WoT platforms. Since cyber-physical systems will be growing in the not-too-distant future, new technologies such as fog and edge computing, Web semantics, machine learning should be considered to create promising WoT platforms for future IoT ecosystems.

\nocite{*}
\bibliographystyle{plainnat}
\bibliography{article}

\end{document}